\documentclass[journal]{IEEEtran}
\usepackage{array, xspace, subfigure, url, xcolor, amsmath, bm, array, graphicx, balance, setspace,tabularx}

\usepackage[linesnumbered,ruled,vlined]{algorithm2e}
\usepackage[noend]{algpseudocode}

\newcommand{\etc}{\emph{etc.}\xspace}
\newcommand{\ie}{\emph{i.e.,}\xspace}
\newcommand{\eg}{\emph{e.g.,}\xspace}
\newcommand{\first}{\textsf{(i)}\xspace}
\newcommand{\second}{\textsf{(ii)}\xspace}
\newcommand{\third}{\textsf{(iii)}\xspace}
\newcommand{\sys}{OpReduce\xspace}

\newcommand{\paraspace}{\vspace{0.01in}}
\newcommand{\parab}[1]{\paraspace\noindent{\bf #1}}

\usepackage{eucal}

\usepackage[font={small,bf},skip=1.0pt]{caption}
\begin{document}
\title{\huge Managing Recurrent Virtual Network Updates in Multi-Tenant Datacenters: A System Perspective}
\author{Zhuotao~Liu,
        Yuan~Cao, and~Xuewu~Zhang
\thanks{Manuscript received Mar. XX, 2018; revised XXX XX, 2018.}
\thanks{Z. Liu is with the NetInfra team of Google Research, USA.}
\thanks{Y. Cao and X. Zhang are with the College of Internet of Things Engineering, Hohai University, Changzhou 213022, China.(e-mail: 20161965@hhu.edu.cn).}}

\markboth{IEEE TRANSACTIONS ON PARALLEL AND DISTRIBUTED SYSTEMS,~Vol.~XX, No.~X,~May~2018}%
{Shell \MakeLowercase{\textit{et al.}}: Bare Demo of IEEEtran.cls for IEEE Journals}

\maketitle
\begin{abstract}
With the advent of software-defined networking, network configuration
through programmable interfaces becomes practical,
leading to various on-demand opportunities for network routing update
in multi-tenant datacenters, where tenants have diverse
requirements on network routings
such as short latency, low path inflation, large bandwidth, high reliability, \etc
Conventional solutions that rely on topology search coupled with an objective function
to find desired routings have at least two shortcomings: \first they run into
scalability issues when handling consistent and frequent routing updates and
\second they restrict the flexibility and capability to satisfy various routing requirements.
To address these issues, this paper proposes a novel search and
optimization decoupled design, which not only
saves considerable topology search costs via search result reuse,
but also avoids possible sub-optimality in greedy routing search algorithms by
making decisions based on the global view of all possible routings.
We implement a prototype of our proposed system, \sys, and perform
extensive evaluations to validate its design goals.
\end{abstract}


\section{Introduction}
Along with the rise of cloud computing, multi-tenant datacenters grow into the
scale of tens thousands of servers hosting millions of tenant VMs~\cite{amazon_dc_size}.
Managing multi-tenancy at such a scale to ensure efficiency,
scalability and agility is a challenging problem drawing considerable 
research and engineering attention. Prior proposals~\cite{netlord, vmware} place
major management effort at hypervisors for the sake of easy configuration and implementation,
whereas little effort has been made to manage \emph{in-network} routings, 
particularly at the granularity of tenants. As a result, 
although the VM locations of each tenant are decided, datacenter network operators 
lose the visibility of actual in-network traffic forwarding for each tenant. 

The lack of tenant-level traffic accountability and routing control could lead
to various limitations. From the perspective of business, 
network operators cannot customize tenant routings in accord with
tenants' service level agreements (SLA) (\eg latency, bandwidth, reliability or security
requirements), which may close the door for such a business model in the virtual private
cloud (VPC) market. From the perspective of management, 
in case of resolving hot spots in datacenters, for instance, it is difficult 
to determine the affected tenants and effectively re-route 
their traffic around congestion in time. 

Thus, it is desirable to have explicit tenant routing control. 
The traditional way to achieve tenant routing management relies on topology search  
coupled with an objective function to greedily find the desired overlay network 
for each tenant. Although instant network configuration is   
technically enabled by SDN~\cite{openflow}, 
the conventional approach still has at least two shortcomings. 
First, tenant routing updates in our production datacenters are recurrent, which can be triggered 
by various reasons including network load dynamics, tenant arrivals/departures, hot spots, 
link failures and so on. Repeatedly performing topology search for each routing update 
will impose significant search cost. Second, a tenant routing is updated typically to 
fulfill certain goals. However, it is uncertain that topology search coupled with 
an objective function is sufficient to achieve any goal. For instance, optimizing a performance metric depending 
on virtual links (\ie VM pair communications) is subject to sub-optimality since 
the mapping between physical links and virtual links is unknown during topology 
search (\S\ref{sec:embedding_goal}).

To address these issues, we propose \sys, a novel search and optimization 
\emph{decoupled} design for routing management. For each tenant-routing update, 
\sys first comprehensively searches all desired overlay candidates
considering only the tenant's VM locations. Then it applies 
a global objective function over these candidates to finalize the most desired one. 
\sys's decoupled model offers at least two advantages. 
First, since topology search is not directed by any objective function, 
search results are general and can be reused across routing updates that 
share the same VM locations: \ie regardless of their goals, topology search 
for future routing updates is saved as long as their VM locations have been explored. 
Second, with the global view of all routing candidates, objective functions 
are not limited to be greedy and local. This eliminates 
the possibility of sub-optimality even when optimizing complex performance metrics, 
for instance, involving virtual links. 

One concern of the decoupled design may be that comprehensively finding 
all desired overlay candidates for a tenant can be expensive. 
However, since the VMs of one tenant are typically
spanning across a few racks, we can reduce the search space to a 
subgraph of the entire topology. Further, most data 
center topologies (\eg VL2~\cite{vl2}, fattree~\cite{fat-tree})
are hierarchically organized into several layers and network traffic 
is never forwarded back and forth between different 
layers. We can adopt these properties to further refine the search space. 
Our proposed algorithm can perform comprehensive 
topology search in polynomial time (\S \ref{sec:search_brief}).

\begin{figure*}[t]
	\centering
	\begin{tabular}{|c|c|c|c|}
		\hline
		\subfigure{\includegraphics[scale=0.2]{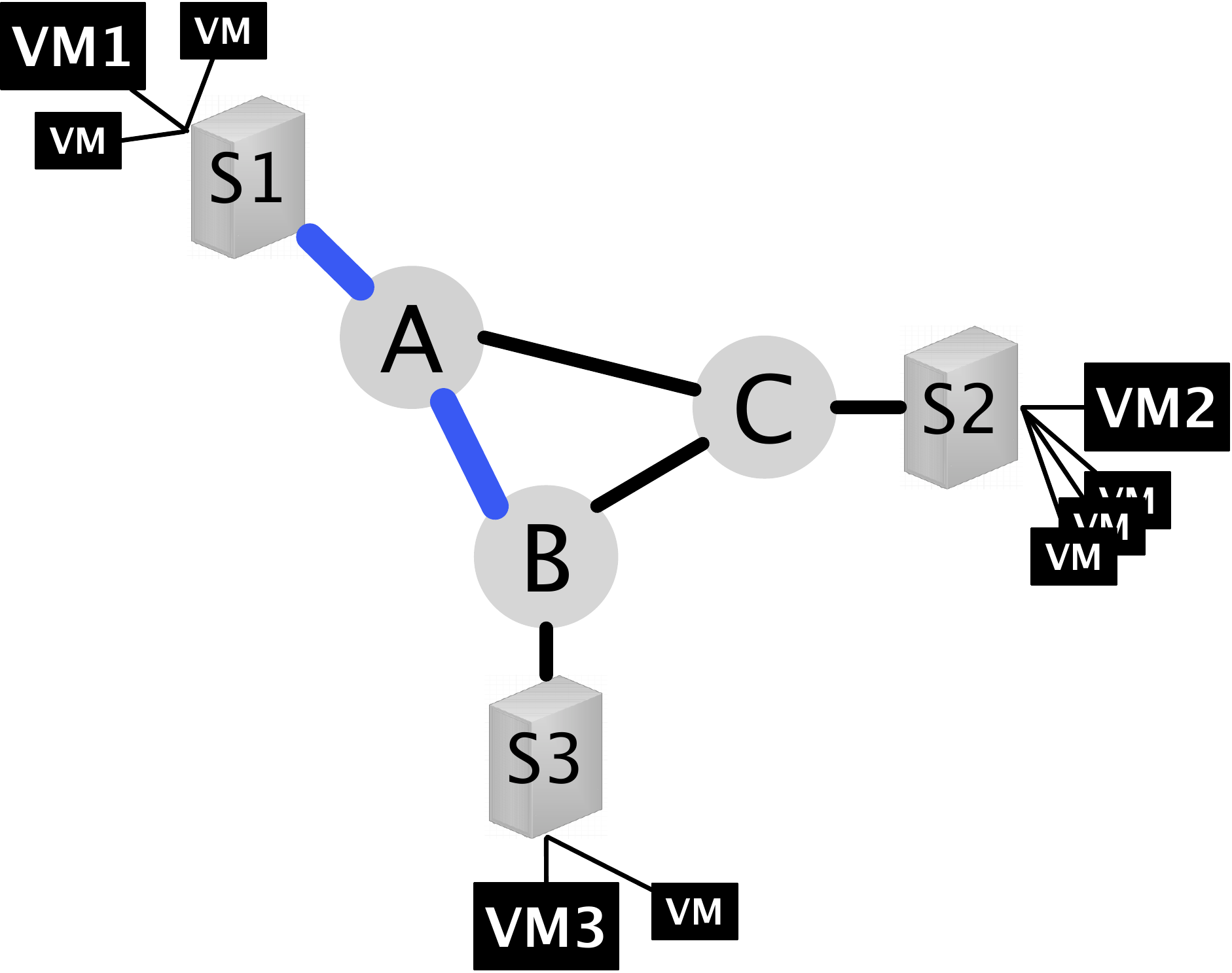}}&
		\subfigure{\includegraphics[scale=0.2]{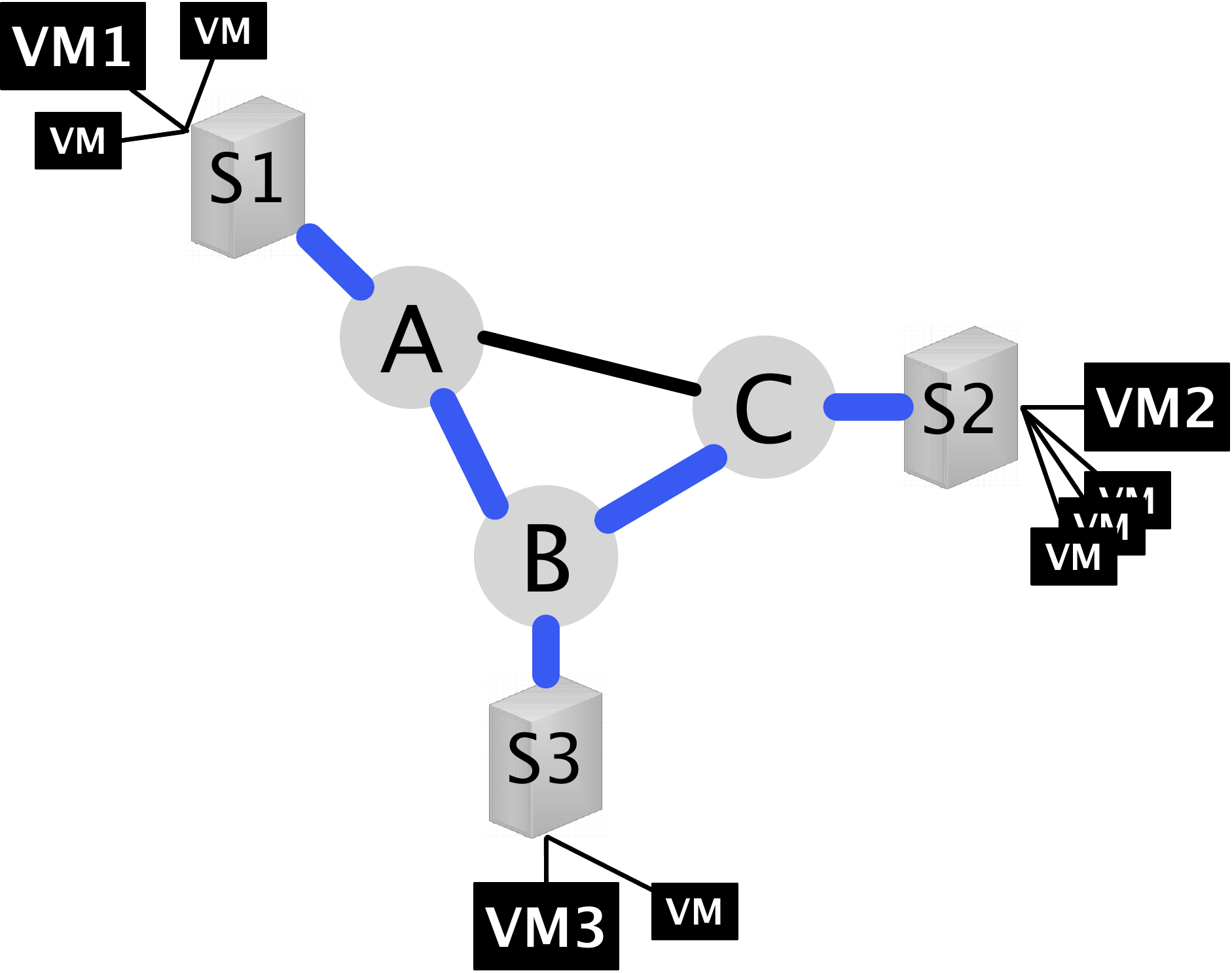}}&
		\subfigure{\includegraphics[scale=0.2]{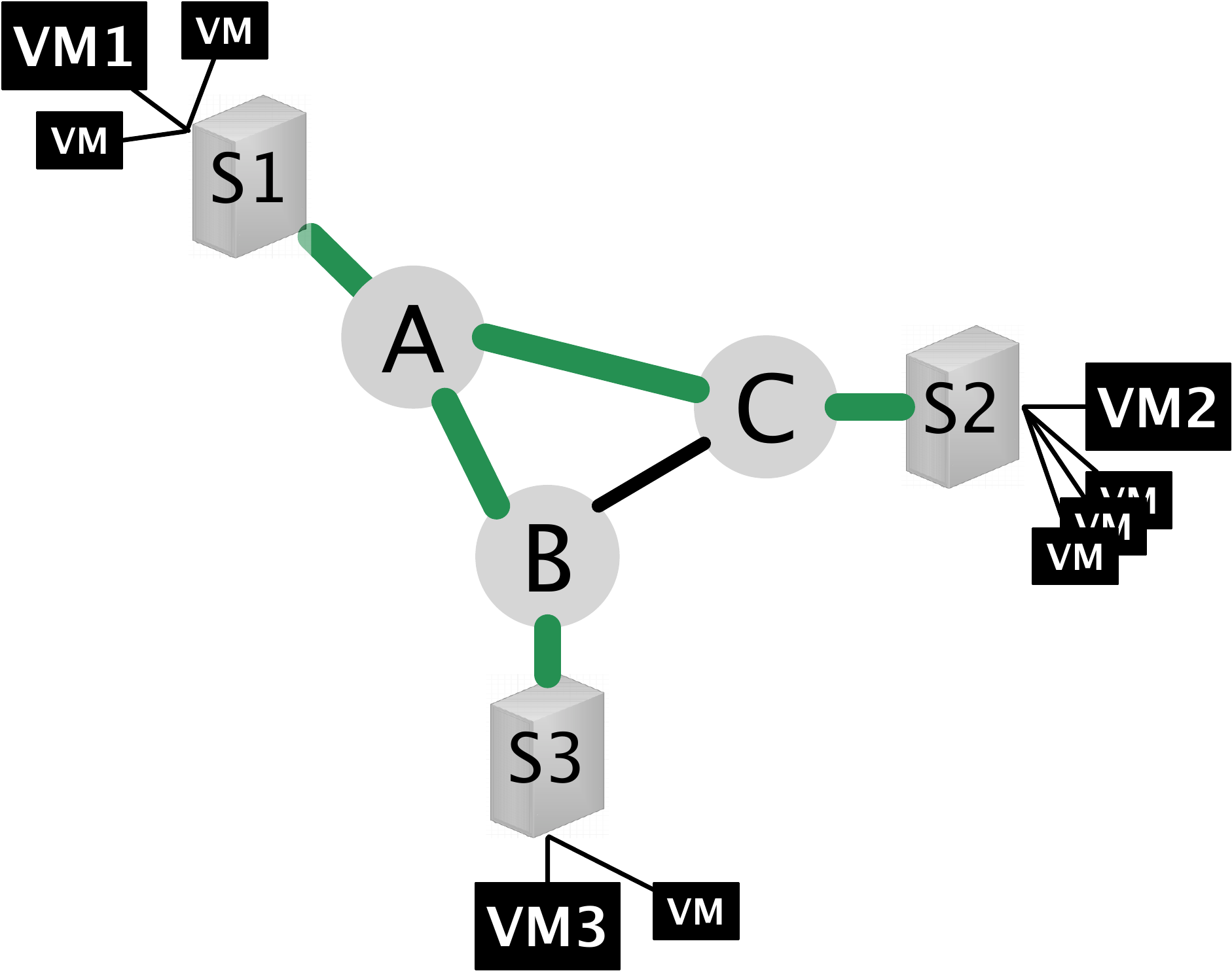}}&
		\subfigure{\includegraphics[scale=0.2]{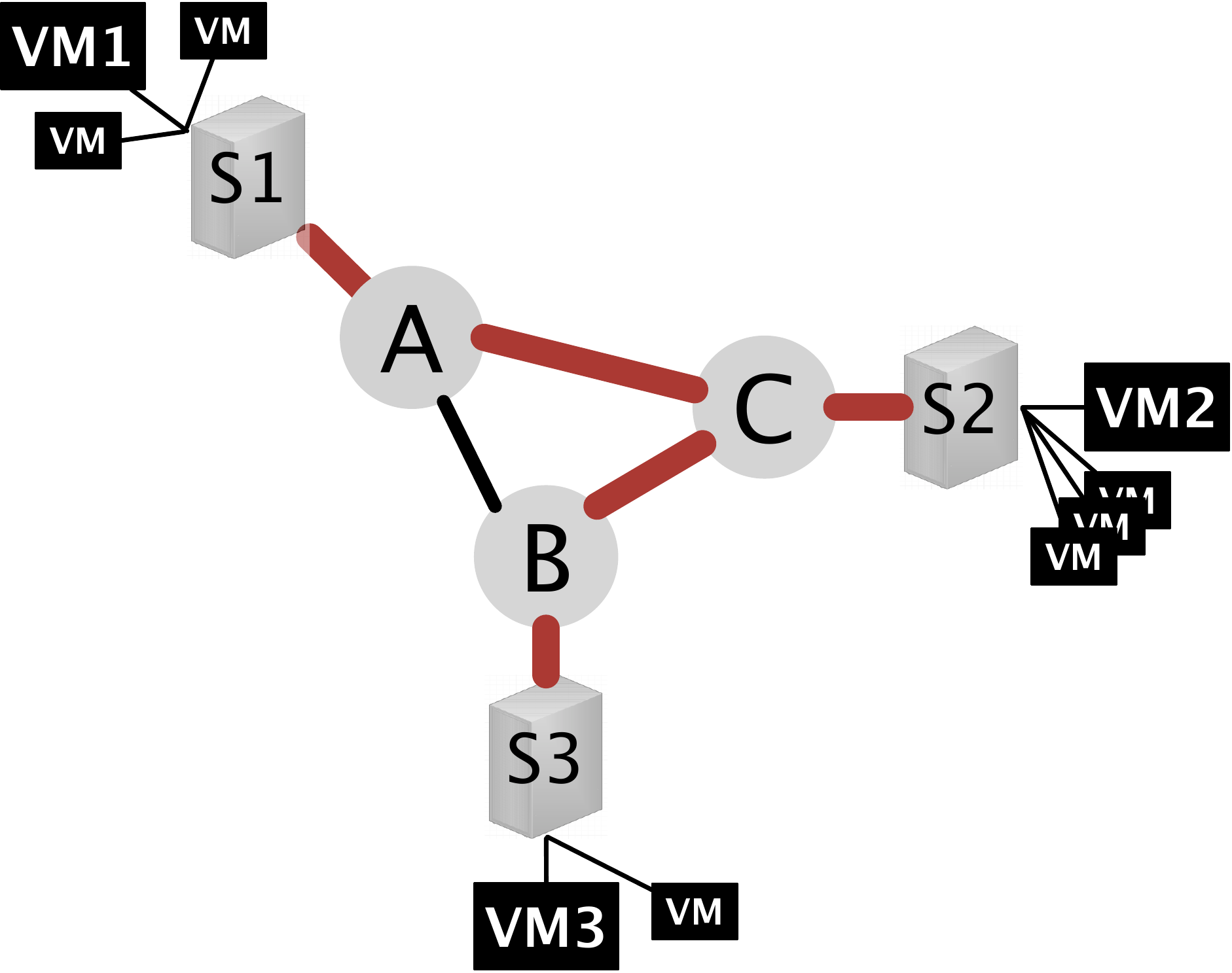}}\\
		\hline
	\end{tabular}
	\begin{tabularx}{\textwidth}{*4{>{\centering\arraybackslash}X}}
		\small (a) Intermediate. &
		\small (b) Default. &
		\small (c) Alternative. &
		\small (d) Optimal. \\
	\end{tabularx}
	\caption{Due to the lack of mapping between virtual links and physical links, 
		conventional search \& optimization coupled solution is subject to sub-optimality when optimizing 
		performance metrics depending on virtual links.}
	\label{fig:mst}
\end{figure*}

We implement a prototype of \sys and perform extensive evaluations 
to validate its design goals. On the one hand, we show that \sys greatly 
reduces the search cost for tenant-routing updates, meanwhile imposing  
small system overhead for managing large scale datacenters (\eg small 
routing cache size and agile network configuration). On the other hand, 
we demonstrate that \sys is able to achieve complex optimization goals 
for routing updates, yielding significant networking performance gain  
over common practice. 

\section{Background and Motivation}\label{sec:background}
\subsection{Managing Per-Tenant Routing}\label{sec:routing_control}


For configuration simplicity, encapsulations protocols such as
VXLAN~\cite{vxlan} and NVGRE~\cite{nvgre} are widely adopted 
in cloud~\cite{vmware}. However, by simply tunneling tenant traffic, 
network operators lose the traffic visibility and accountability, \ie they 
are not able to identify the origin of network traffic since packets are sent 
on behalf of hypervisors and one hypervisor may host VMs for multiple tenants. 
Such invisibility becomes even worse when datacenters run link aggregation
and/or perform load balancing (\ie ECMP, Hedera~\cite{hedera}, or CONGA~\cite{conga}) 
to spread traffic across redundant physical links. As a result, even if
one tenant occupies only a small fraction of computation resources 
in the datacenter, 
its traffic could appear on many physical links, making network operators 
unaware of actual in-network traffic forwarding for the tenant. 

The lack of in-network tenant-routing management causes various
operation restrictions. For instance, it is difficult for network operators to
perform monitoring, measurement and trouble-shooting for a tenant as
its traffic may spread across many network links.
Further, as virtual private cloud gains popularity, more tenants have 
incentives to customize their private cloud based on their own needs, \eg 
one tenant may want to promote latency performance of web servers,  
whereas someone else may want to optimize bandwidth for
MapReduce tasks. Without explicit tenant-level 
routing control, it is difficult to customize routings for
individual tenants.

Thus, it is desirable for network operators to have explicit control over tenant routings. In particular, for each
tenant, network operators explicitly configure a Virtual Tenant Network
(VTN), which is an overlay network connecting
the tenant's VMs. The tenant's traffic is confined within its VTN and no other physical links 
besides the ones in its VTN can carry the tenant's traffic. 
As tenants can be identified by their VTNs, traffic is accountable, 
which eliminates these management limitations. 
Further, network operators can customize a tenant's routing based on its 
requirement by embedding its VTN into an overlay that can best satisfy the 
requirement. 

The traditional way of embedding VTN for a tenant relies on topology search 
coupled with an objective function to find the desired overlay. 
Such an approach, however, has at least two shortcomings: \first it runs into scalability 
issues when handling frequent VTN embedding requests and \second it may 
be insufficient to find the optimal overlay for complex embedding goals.
Next, we elaborate on the two issues. 

\subsection{Frequent VTN Updating Requests}\label{sec:all_time}
In production multi-tenant datacenters, managing VTN embeddings/updates are all-time tasks on the basis
of individual tenants, which are up to tens of thousands~\cite{amazon_dc_size}. 
VTN embedding requests are triggered by many sources including 
network load dynamics, link congestions and failures, hot spots, 
tenant departures and so on. 
Although efficient graph search algorithms, such as Prim's and
Kruskal's, have been proposed for decades, consistently and frequently
performing overlay search in large topology for a large number of VTN updates still 
imposes significant cost. Resolving the scalability problem 
is the first step towards efficient tenant-routing management.

\subsection{Fulfilling VTN Embedding Goals}\label{sec:embedding_goal}
A VTN embedding/update is as simple as finding a random overlay 
network for the tenant. Instead, the VTN may need to satisfy various 
requirements such as guaranteed bandwidth~\cite{predictable,secondnet,tag},  
bounded latency~\cite{silo,qjump}, required security appliances
~\cite{cloudwatcher,netsecvisor}, and other performance metrics 
depending on the tenant's SLA. As these embedding goals become more complex, 
it is uncertain that the conventional search \& optimization coupled solution can find 
the real optimal routing. 

We use an illustrative example in Figure~\ref{fig:mst} to demonstrate such sub-optimality. 
Network operators perform VTN update for a tenant whose VMs are hosted by the set of 
hypervisors $[\textsf{S}_\textsf{1}, \textsf{S}_\textsf{2}, \textsf{S}_\textsf{3}]$. 
The embedding goal is to minimize the communication cost of all virtual links. 
For simplicity, we define the communication cost of a virtual link as  
the number of hops on the path connecting the two VMs of the virtual link. 
Thus, if VM1 and VM2 take path 
$\textsf{S}_\textsf{1}\textsf{-}\textsf{A}\textsf{-}\textsf{C}\textsf{-}\textsf{S}_\textsf{2}$,
the cost is 3. 

This embedding is a classic minimum spanning tree problem that can  
be solved by a greedy topology search algorithm. 
At certain intermediate state of the overlay search (Figure \ref{fig:mst}(a)), 
\textsf{AB} has been appended to the overlay. Based on the tie breaker in the algorithm, 
\textsf{BC} is then appended to the ongoing
overlay, and the final overlay returned by the algorithm is $\textsf{S}_\textsf{1}\textsf{-A}$, 
$\textsf{S}_\textsf{2}\textsf{-C}$, $\textsf{S}_\textsf{3}\textsf{-B}$, 
\textsf{AB}, \textsf{BC} (Figure \ref{fig:mst}(b)). With a different algorithm, 
\textsf{AC} could be the link to append after state (a) rather than \textsf{BC}, and the 
final overlay returned will the one shown in Figure \ref{fig:mst}(c). 
However, no matter how the search algorithm is designed, 
the optimality of an overlay is up to the distribution of VMs. 
Although VM locations are known a priori, in an intermediate state, 
the search algorithm cannot determine exactly which virtual links a physical link 
will carry. Thus, the search algorithm cannot guarantee to capture the real optimal 
overlay (Figure \ref{fig:mst}(d)). In fact, due to the lack of mapping between virtual links and physical links, 
the conventional search \& optimization coupled solution is subject to  
sub-optimality when optimizing any performance metric involving virtual links. 

\emph{We note that providing rigid theoretical proof to show the sub-optimality of 
	 the conventional search \& optimization approach is beyond the scope 
	 of this paper. Instead, this paper focuses on offering a system perspective 
	 of how tenant routings are managed in practice, especially based on our experience with 
	 production datacenters.}

\subsection{Decoupling Search and Optimization}\label{sec:decouple}
To address the above challenges, \sys proposes a search and optimization 
decoupled design. For a routing update request, \sys first comprehensively finds all 
desired overlay candidates considering only VM locations. Then it applies 
an objective function, designed with the global view on all routing candidates, 
to finalize the optimal one. In the above example, \sys will first obtain all 3 
overlay candidates (figures (b), (c) and (d) in Figure~\ref{fig:mst}) and 
then evaluate them to find the optimal one (Figure \ref{fig:mst}(d)). With the global view 
on all candidates, \sys knows the exact mapping between virtual links 
and physical links. Thus, it can design a global objective function in evaluation to 
guarantee optimality. Further, as topology search is not directed by any objective 
function, search results are general so that they can be reused for further VTN updates that 
have the same VM locations. This saves considerable cost for performing 
frequent VTN updates in large scale datacenters.

\section{System Design} \label{sec:system}
\begin{figure}[t]
  \centering
  \mbox{
    \subfigure{\includegraphics[scale=0.4]{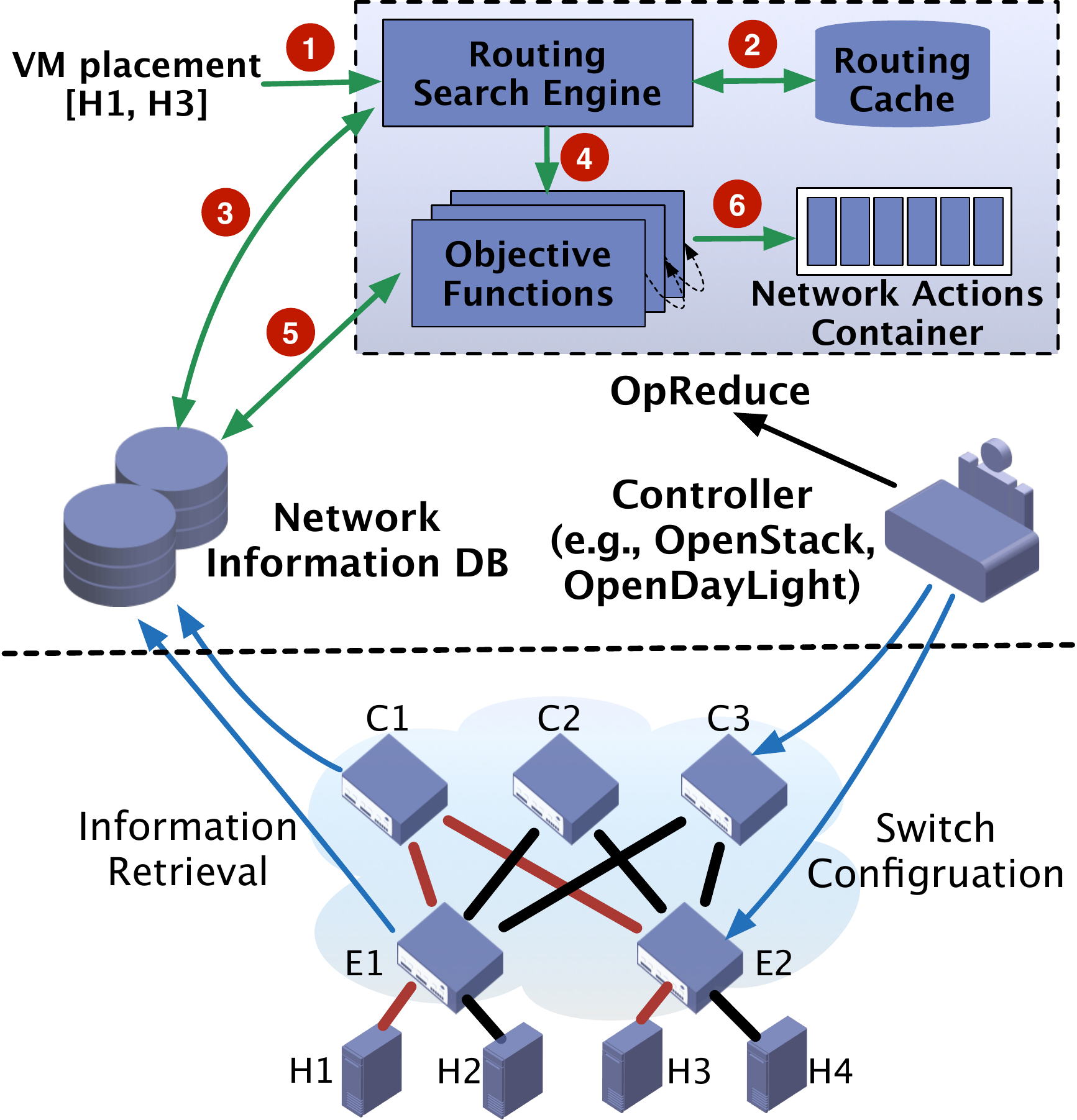}}
    }
  \caption{The architecture of \sys.}
  \label{fig:system_arch}
\end{figure}

In this section, we elaborate on the design of \sys. 
Figure \ref{fig:system_arch} illustrates the architecture of \sys. 
\sys is built upon a network information 
database and a controller. The network information database 
allows \sys to retrieve network related information, such as network topology and link 
utilization. The controller is used to manage both computation and networking resources, such as 
assigning VMs to tenants and configuring VTNs for tenants. 
\sys's decoupled design is achieved by its VTN embedding module which includes 
routing search engine, routing cache, objective functions and network action container. 

We now describe the workflow of performing routing update using \sys. 
Upon receiving a request, \sys first obtains a list 
of desired routing candidates based on the VM placement. These routing candidates may 
come from either the routing search engine or the routing cache if the VM placement has been 
explored before. Then \sys evaluates each candidate to determine the most desired one 
for achieving a specific goal. To assist evaluation, network operators propose 
evaluation methods, defined as  objective functions, to score each routing candidate. 
The design of objective functions is facilitated  by the global view of all possible routing candidates. 
Further, useful network information (\ie link loads) is retrieved from the network information database. 
Finally, after determining the most desired routing, \sys enforces the VTN embedding inside the network 
by executing network tasks (\ie switch configuration) via the network action container. 

\emph{Although setting up both the 
network information database and the network/computation controller takes considerable 
implementation efforts, we omit these details to focus on the research part of \sys in this paper.} 
Next, we describe each individual component of \sys's VTN embedding module. 
Hereafter, we use \emph{routing} and \emph{VTN} interchangeably to indicate the overlay 
network carrying a tenant's traffic.

\normalsize

\subsection{Routing Search Engine}\label{sec:search_brief}

Given a certain VM placement, the routing search engine is used to produce a list of \emph{desired} routings 
(or routing candidates). In the typical layered datacenter network layout (\eg 
fattree~\cite{fat-tree}, Clos~\cite{vl2,clos}), a routing is desired if traffic between two VMs does not 
bounce back and forth more than once between two different layers. For random topology (\eg~\cite{jellyfish,SWDC}), 
we can bound the maximum number of hops on VM-pair paths 
to exclude undesired routings. As datacenter fabric is often built with a high level of path redundancy, 
exploring all desired routings will provide sufficient candidates to achieve the embedding goal 
of the routing update request.

Next, we propose an algorithm that can produce a comprehensive list 
of all desired routings in layered data center topology. In the network, we associate each node 
with a \emph{height}. All core switches are assigned height 
$1$ and the initial height for other nodes is infinity. Then we trace down from core switches 
towards other nodes to assign them a height. The height of 
a node is the minimum height among all its neighbours plus one. 
All nodes with the same height compose a layer whose layer number is the node height.
We define a path as \emph{straight} if   
all nodes on the path have different heights. 
Straight path is either upward (from a source at a lower layer to a 
destination at a higher layer) or downward. 

In a desired routing, to connect two hypervisors $h_1$ and $h_3$, 
$h_1$ needs to have a straight path which shares the same endpoint with one of 
$h_3$'s straight paths. 
The shared endpoint is defined as a \emph{common node} for the hypervisor set
$[h_1,h_3]$. Formally, a node is defined as the common node for a set of hypervisors 
if it has at least one straight path to reach each hypervisor in the set. 
Then traffic from $h_1$ can first take one straight path to reach the common node and 
then bounces back to reach $h_3$ via another straight path. 
Thus, a routing connecting 
$[h_1,h_3]$ is composed of the above two straight paths. Similarly, a routing 
connecting all VMs of a tenant is composed 
of several straight paths, among which each straight path is 
originated from one hypervisor and all these paths are ended at 
one same common node. 

We formulate our search procedure in Algorithm~\ref{algo:search}. 
At the very high level, the algorithm works as follows: \first find all common nodes for the 
hypervisor set (line~\ref{line:common_nodes}); \second for each common node, find one downward straight path 
from the common node to each hypervisor (line~\ref{line:downwardpath}) 
and \third combine these paths to produce one routing candidate (line~\ref{line:combine}). 
Since a routing candidate can only contain one common node (having more will produce loops), 
Algorithm~\ref{algo:search} provably finds all desired routings in the network. 
Even though the algorithm comprehensively finds all desired routings, the algorithm complexity 
is polynomial (not exponential). We defer detailed complexity analysis in \S \ref{sec:appendix}.

\newcommand{\algrule}[1][.2pt]{\par\vskip.5\baselineskip\hrule height #1\par\vskip.5\baselineskip}
\begin{algorithm}[t]
\SetKwData{GUP}{GetUpwardGraph}
\SetKwData{GetDownwardPath}{GetDownwardPath}
\SetKwData{hs}{hypervisor set}
\SetKwData{cm}{CommonNodes}
\SetKwData{func}{\textbf{Function:}}
\SetKwData{IP}{\textbf{Input:}}
\SetKwData{OP}{\textbf{Output:}}
\footnotesize

\IP VM placement $H$. \\
\OP All desired routing candidates.\\
\algrule[0.5pt]

	\For{$h_i \in $ \hs $H$}{
		$\mathcal{T}_i \leftarrow$ \GUP{$h_i$}; ~~~$\mathcal{T} \leftarrow \bigcup_{i} \mathcal{T}_i$\; \label{line:upwardgraph}
	}
	\cm $\leftarrow \bigcap_{\mathcal{T}_i \in \mathcal{T}} \mathcal{T}_i$.nodes()\; \label{line:common_nodes}
	\ForEach{$c_j \in$ \cm}{
		\For{$h_i \in$ \hs $H$}{
			$\mathcal{N}_j \leftarrow \bigcup_{i}$ \GetDownwardPath{$c_j,~h_i,~\mathcal{T}_j$}\; \label{line:downwardpath}
		}
	}
	\Return $\mathcal{N} \leftarrow \bigcup_{i} \mathcal{N}_i$\;\label{line:combine}
	
\algrule[0.5pt]

\func \GUP($h_i$) \\ 
\quad \Return The graph containing all upward straight paths starting from $h_i$ and ending 
at core switches.  

\algrule[0.8pt] 

\textbf{\func} \GetDownwardPath{$c_j,~h_i,~\mathcal{T}_j$} \\  
\quad \Return The straight path from $c_j$ to $h_i$ in graph $\mathcal{T}_j$.

\caption{\bf Routing Search Algorithm}\label{algo:search}
\end{algorithm}
\normalsize

For a random data center topology~\cite{jellyfish,SWDC}, \sys can first adopt the $k$-shortest path
algorithm~\cite{shortest} to obtain a set of paths between each hypervisor pair 
and then combine these paths to produce routing candidates. 
$k$ can be parameterized to exclude undesired routings. 

\subsection{Tenant-Routing Cache}
The search results for routing candidates are cached using a dictionary (hash table) data
structure, where the key is VM placement 
and the value is a list of all desired routings for the VM placement. 
Each routing is stored as a list of physical link IDs, which can be used to retrieve 
link related information (\eg utilization, status) from the network information database. 
Generally, entries in the dictionary are valid as long as the 
network topology remains the same. In \sys's prototype, 
we associate each entry in the dictionary with a relatively long validation 
period (\eg few weeks), and re-perform routing search after 
an entry is expired. 

In \sys's implementation, we allocate $32$ bits for both the hash key and physical link IDs. 
In \S \ref{sec:evaluation:cache_size}, we show that even in large scale $k{=}32$ fat-tree datacenter 
with over $10$ thousands servers (millions of VMs), the cache size for managing $10$ thousand tenants is 
about $100$ MB, which can be easily managed by commodity servers. 

\subsection{Objective Functions}
The objective functions are used to evaluate routing candidates so as to determine the 
most desired one for fulfilling embedding goals. For instance, a valid
objective function can be as sophisticated as a combination function balancing
latency, bandwidth, the number of hops and so on. Or it can 
be as detailed as optimizing specific virtual links for latency and other virtual 
links for bandwidth. In general, \sys is open to accept any objective function. 
But \sys offers global views on all possible routing candidates so that objective 
functions are not limited to be greedy and local.  Relevant network information, available 
in \sys's network information database, can be applied to evaluate these routing candidates. 
By the time of this writing, an instance of \sys deployed in our production datacenter has 
pre-installed tens of abstract objective functions that can be parameterized,  
allowing network operators to achieve a wide variety of VTN embedding goals.

\subsection{Network Action Container}
After determining the most desired routing, the final step is enforcing the routing 
inside network. The enforcement process is essentially configuring switches on the 
routing (\eg configuring VLAN tags~\cite{vlan} or adding OpenFlow rules~\cite{openflow_spec}) 
to guide the switches to perform desired traffic forwarding.   
To facilitate routing management in large scale datacenters, 
we develop a network action container to perform network configurations in a batch. 
For instance, configurations on different ports of one switch are aggregated in 
one thread and configurations on different switches are paralleled via multi-threading. 
As shown in \S \ref{sec:evaluation:switch_config}, the network action container significantly reduces 
the overall configuration latency. 

\section{Implementation}\label{sec:implementation}

We have a full implementation of \sys. We use OpenStack to create VMs on hypervisors and assign VMs to 
tenants. On our testbed dedicated for experiments, the OpenStack environment 
has $4$ compute nodes as hypervisors that can host about one hundred VMs, 
one network node, and one controller node. Because OpenStack is limited to computation 
virtualization, we unify it with our physical network via an 
OpenDayLight~\cite{opendaylight} controller that uses OpenFlow and \textsf{netconf}~\cite{netconf}
protocols to manage OpenFlow and legacy switches, respectively.

On our testbed, we allocate a dedicate VLAN tag for each
VTN. Then embedding a VTN is about configuring relevant switch interfaces
with corresponding VLAN tags to achieve reachability. Certainly there could be other 
network virtualization solutions, \eg slices in sliceable
switch~\cite{trema}, VLAN tag stacking~\cite{stacked_vlan}. \sys's 
network controller is extendible to support these virtualization solutions. 
We are aware that VLAN tags are limited to $4096$ values 
which may be insufficient in large scale datacenters. To resolve this scalability issue, 
\sys implements a helper component, based on Panopticon~\cite{panopticon}, 
to achieve VLAN tag reuse in hybrid datacenters with both legacy and OpenFlow switches. 

\section{Evaluation}\label{sec:evaluation}
Our evaluation centers around the followings. \first In \S \ref{sec:congestion_aware}, we show  
\sys is guaranteed to find the least congested routings for VTN embeddings under various 
settings, which yields significant networking performance 
improvement over common practice. \second In \S \ref{sec:system_evaluation}, 
we show that \sys greatly reduces the search cost for managing numerous routing updates 
and imposes small system overhead while managing routing updates in large scale datacenters.

\subsection{Congestion-Aware Routing Updates}\label{sec:congestion_aware}
One representative goal for routing updates in multi-tenant datacenters is to find 
the least congested routing to re-accommodate a tenant. The traditional search \& 
optimization coupled solution for finding the desired routing is using the 
Prim's minimal spanning tree algorithm with physical link utilization as the edge weight. 
However, since network traffic is generated by 
VM pairs, minimizing the congestion experienced by virtual 
links is the more direct and therefore more accurate goal for finding the least congested routing. 
Thus, \sys uses the following algorithm to determine the most desired routing. 
Assume \sys produces $m$ routing candidates $\{\mathcal{T}_1,...,\mathcal{T}_m\}$ for 
a routing update request that has $n$ virtual links $\{l_1,...,l_n\}$. 
Then the most desired routing is 

\begin{equation}\label{eqn:routing}
\mathcal{T}^* = \min_{\mathcal{T}_k} \frac{\sum_{i=1}^n ~ \lambda_i\cdot \mathcal{F}_k(l_i)}{n},
\end{equation}
where $k\in[1,m]$, $\mathcal{F}_k(l_i)$ is the congestion level experienced by virtual link $l_i$ 
in routing candidate $\mathcal{T}_k$ and $\lambda_i$ is the weight of $l_i$. 
In \sys's instance deployed in our production datacenters, both $\mathcal{F}$ and $\lambda_i$ 
are configurable to enable highly flexible routing customization. 
For evaluation purpose in this paper, $\mathcal{F}_k(l_i)$ is 
defined as the highest congestion level of all physical links in $\mathcal{T}_k$ that carry $l_i$, 
where the congestion level of a physical link is estimated as its average link utilization. 
Further, all virtual links are equally weighted. 

Besides the conventional search \& optimization coupled solution (referred to as the \emph{local} solution), 
we also compare \sys with flow-level ECMP, the common practice for load balancing 
and congestion reduction in datacenters, and the \emph{bottomline} solution which 
embeds VTN to a randomly selected routing candidate. 

For each embedding request, we compare the routing determined by \sys and the ones
selected by the other three solutions. We use average flow completion time (FCT) 
as the metric to quantify the congestion experienced by virtual links
(similar to~\cite{why_fct, conga}), expecting that a better routing 
will have shorter average FCT. \emph{Note that we do not claim that FCT is either the only or 
the optimal metric. We use FCT for the sake of quantifiable results.}

\begin{figure}[t]
	\centering
	\mbox{
		\subfigure[\label{fig:flow:a}\small Enterprise workload\cite{conga}]{\includegraphics[scale=0.35]{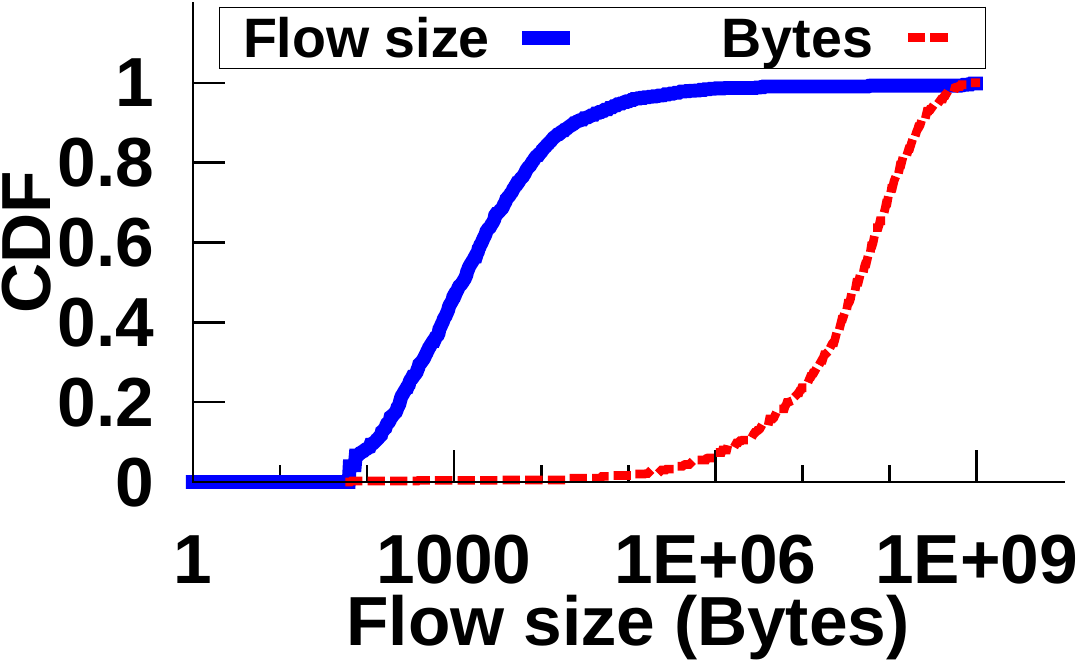}}  
		\subfigure[\label{fig:flow:b}\small Data-mining workload\cite{vl2}]{\includegraphics[scale=0.35]{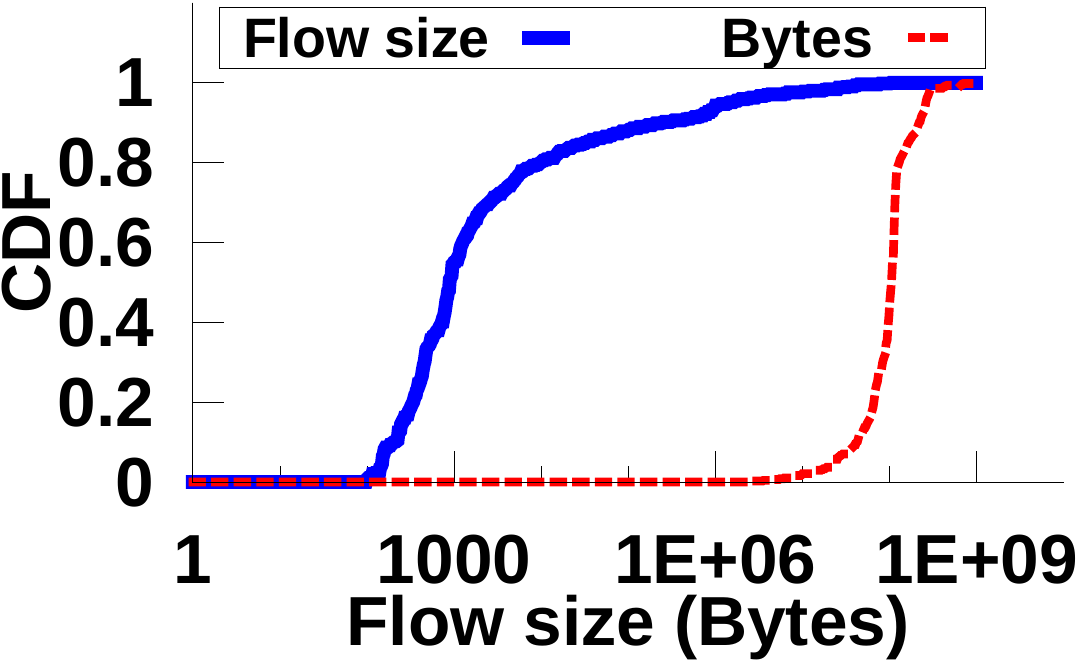}} 
	}
	\caption{Empirical workloads used in evaluation.
	}
	\label{fig:flow}
\end{figure}

\begin{figure}[t]
	\centering
	\mbox{
		\subfigure[\small Background utilization in one snapshot\label{fig:fct_back}]{\includegraphics[scale=0.26]{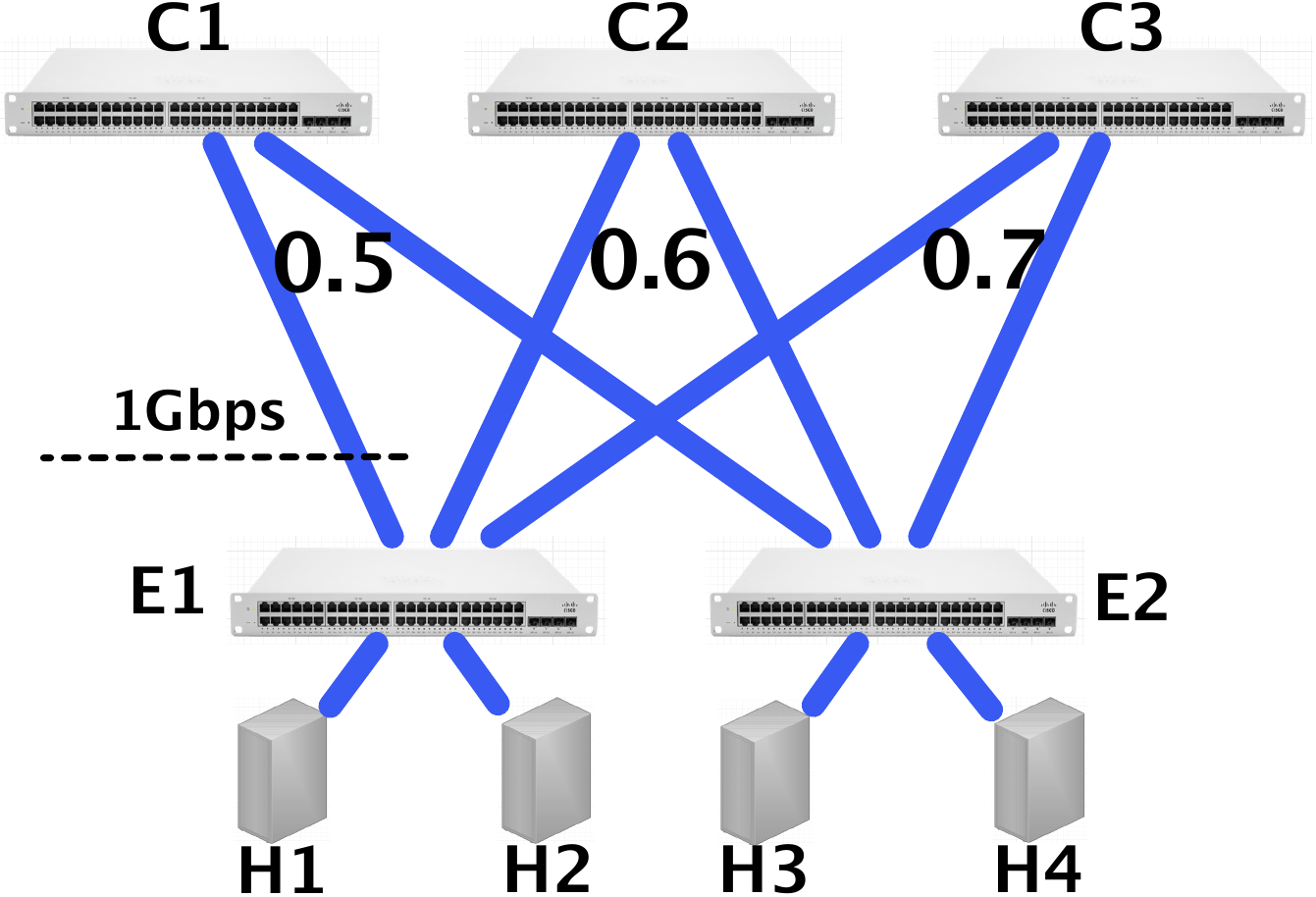}}
		\subfigure[\small Desired routings produced by each solution\label{fig:fct_one_trial}]{\includegraphics[scale=0.26]{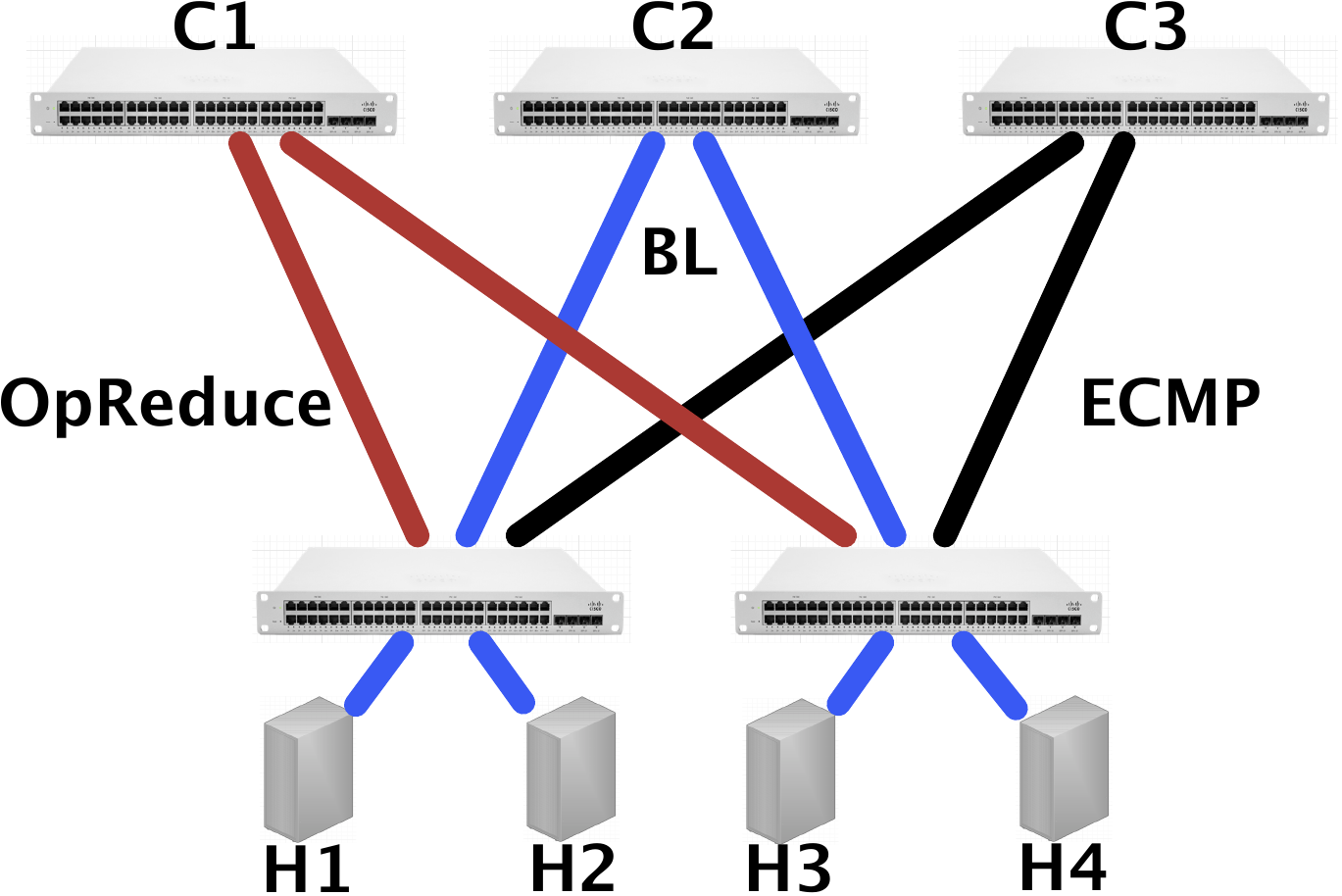}}
	}
	\mbox{
		\subfigure[\small Average FCT in each experiment set \label{fig:fct_all_trial}]{\includegraphics[scale=0.42]{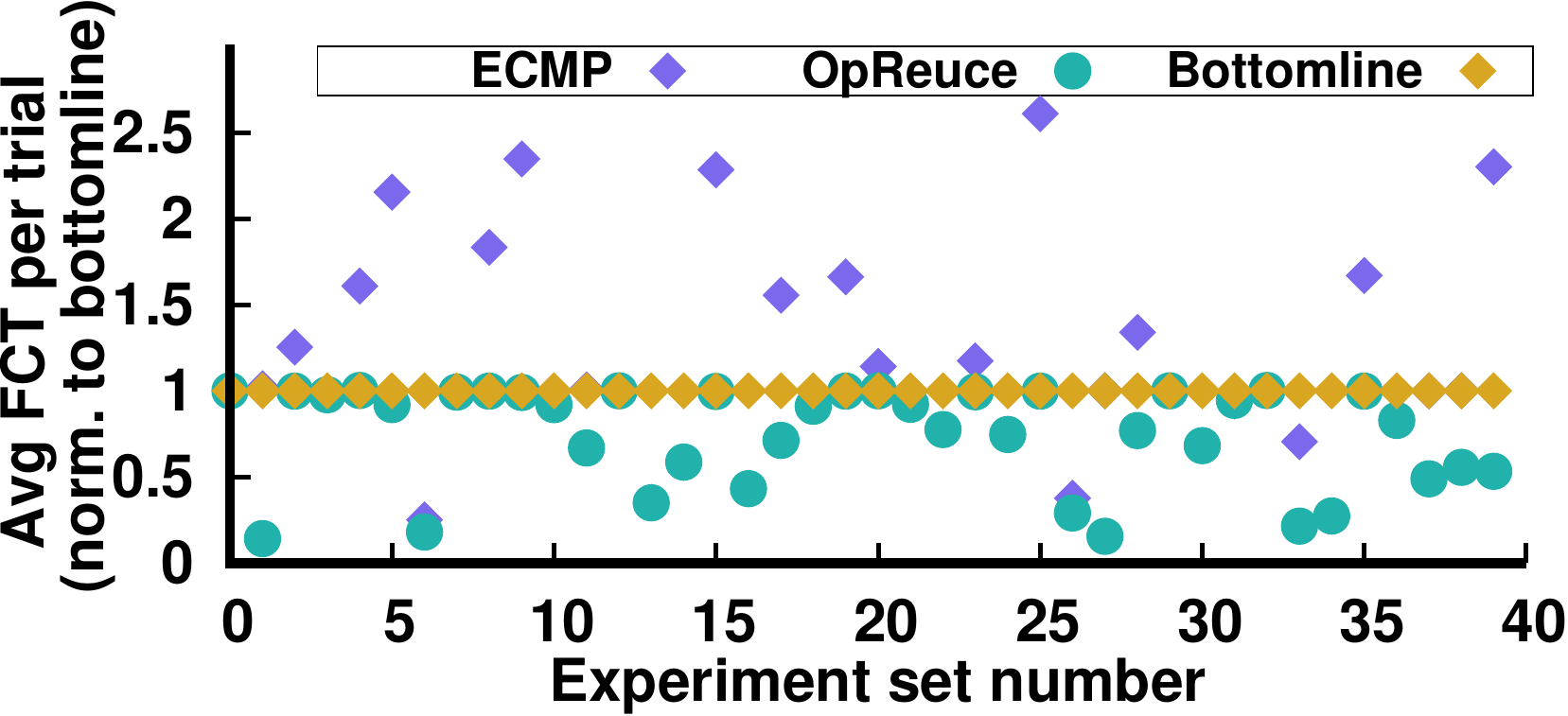}}
	}
	\caption{Testbed experiments for improving FCTs.}
	\label{fig:testbed_CA}
\end{figure}

\begin{figure*}[t]
	\centering
	\mbox{
		\subfigure[\small FCT min-mean-max distribution.\label{fig:baseline:a}]{\includegraphics[scale=0.48]{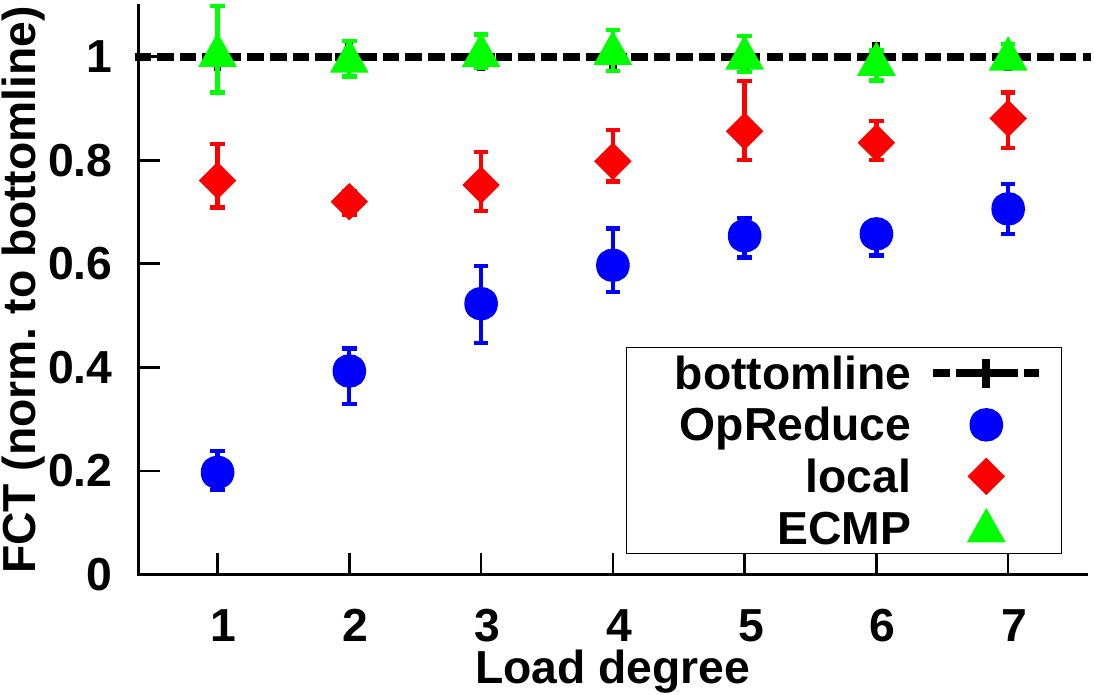}}
		\subfigure[\small False rate.\label{fig:baseline:b}]{\includegraphics[scale=0.48]{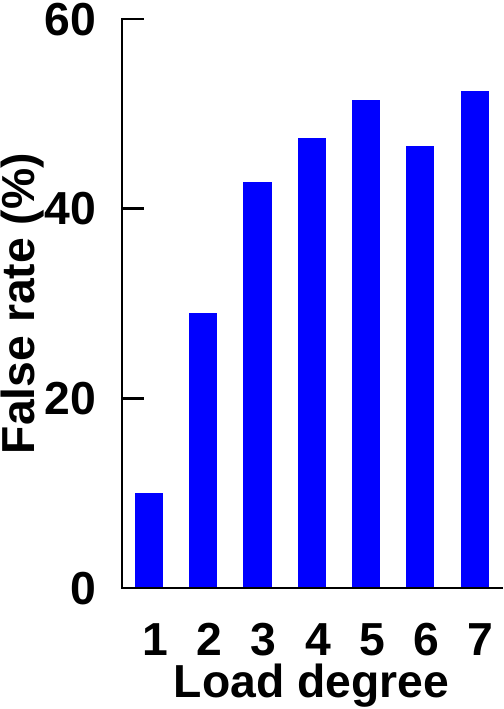}}\quad
		\subfigure[\small FCT min-mean-max distribution.\label{fig:baseline:c}]{\includegraphics[scale=0.48]{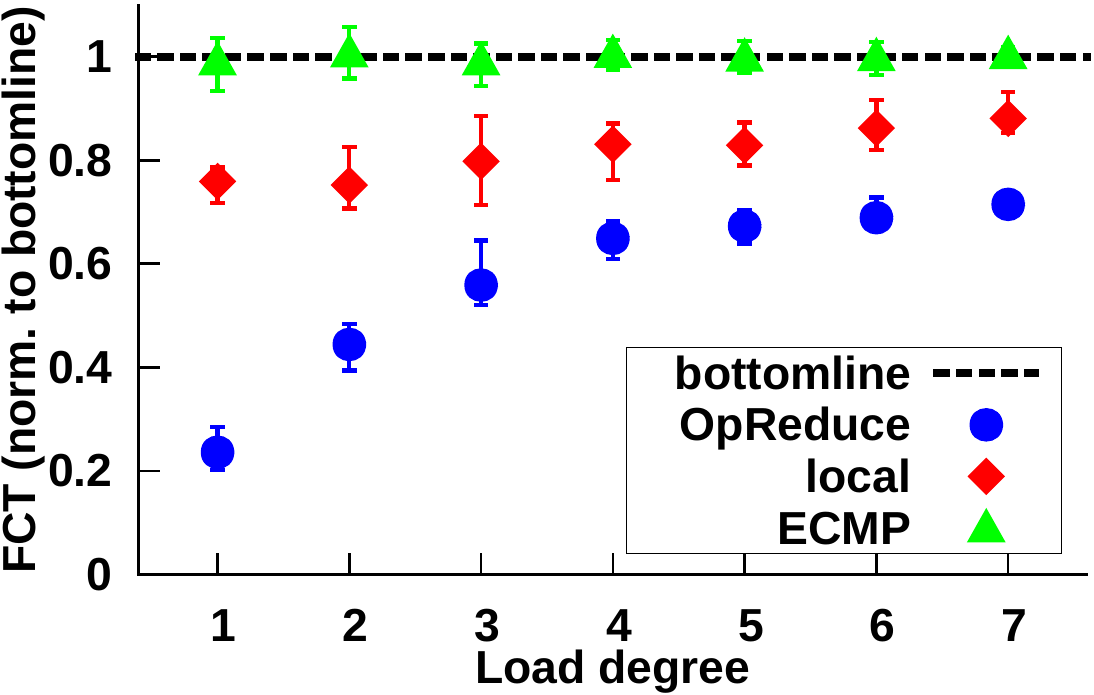}}
		\subfigure[\small False rate.\label{fig:baseline:d}]{\includegraphics[scale=0.48]{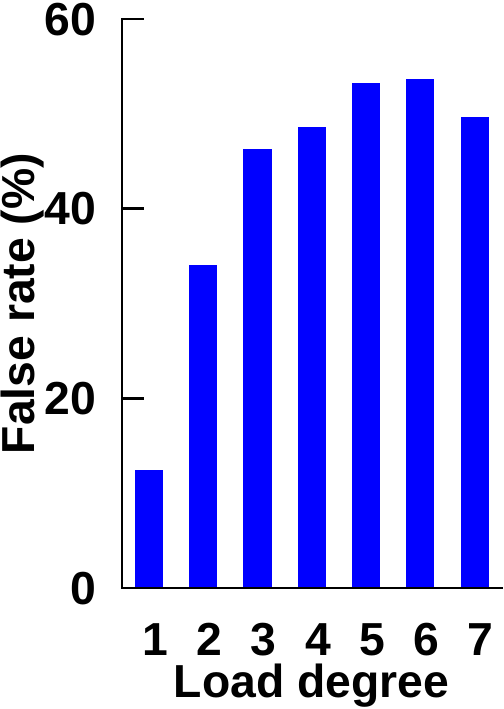}}
	}
	\caption{Impact of the fabric load on \sys's performance.
		The normalized FCT is used as the performance metric. 
		Part (a) and (b) use the enterprise workload~\cite{conga}. Part (c) and (d) use the data-mining workload~\cite{vl2}.
	}
	\label{fig:baseline}
\end{figure*}

\subsubsection{Testbed Experiments}
We start the evaluation on our physical testbed. 
We build a network topology illustrated in Figure \ref{fig:fct_back}. 
The datacenter has pre-embedded tenants to generate traffic 
using our client/server programs. The clients initiate long-lived TCP connections to 
randomly selected servers to request flow transfers.  All VMs run both the client and 
server programs. Only intra-tenant communication is allowed. 

In this experiment, we perform routing update for a tenant who has 10 VMs 
that are randomly distributed in all four hypervisors on our testbed. 
In total, we perform $40$ sets of experiments under different network utilization. 
Figure \ref{fig:fct_back} shows the snapshot of network utilization 
in one experiment set. In each experiment set, we perform $3$ individual routing updates 
using \sys, ECMP and the bottomline solution. 
After the tenant VTN is embedded, its virtual links (VM pairs) start to 
generate traffic using our client/server programs. Flow sizes are randomly 
sampled from the empirical datacenter workload~\cite{conga}. To 
ensure fair comparison, the set of flow sizes used for the 3 experiments in each  
experiment set is identical. We compute the average FCT 
among all flow transfers as our performance metric. 

Figure \ref{fig:fct_one_trial} illustrates the desired routing produced by each solution 
for the network utilization shown in Figure~\ref{fig:fct_back}. 
Since \sys first outputs all routing candidates (\ie 
$\textsf{E}_\textsf{1}\textsf{-}\textsf{C}_\textsf{1}\textsf{-}\textsf{E}_\textsf{2}$, 
$\textsf{E}_\textsf{1}\textsf{-}\textsf{C}_\textsf{2}\textsf{-}\textsf{E}_\textsf{2}$, 
$\textsf{E}_\textsf{1}\textsf{-}\textsf{C}_\textsf{3}\textsf{-}\textsf{E}_\textsf{2}$) and 
then applies the objective function (Equation~(\ref{eqn:routing})) to determine the most desired one, it can 
always find the least congested routing ($\textsf{E}_\textsf{1}\textsf{-}\textsf{C}_\textsf{1}\textsf{-}\textsf{E}_\textsf{2}$ 
in this case). However, both ECMP and bottomline are unaware of the link utilization. 
Consequently, their static hashing could result in overwhelming these more congested links.  

Figure \ref{fig:fct_all_trial} plots the average FCT 
in each experiment set when using enterprise workload~\cite{conga} 
(results for using data-mining workload~\cite{vl2} are similar, and we omit them for brevity).  
Although the bottomline solution and ECMP may produce different FCTs in one set, 
their average FCTs over all sets are close. 
Comparing with these two solutions, \sys can consistently find the 
optimal routing, yielding ${\sim}30$\% less FCTs on average and up to 
${\sim}5$x reduced FCTs in some sets (outside the plot scope of Figure \ref{fig:fct_all_trial}). 

Note we do not compare \sys with the local solution in our testbed experiments. This is because 
the local solution and \sys will produce the same routing in such a small topology. Next 
we show that given large network topologies, it is very likely for the local solution 
to return sub-optimal overlays, which results in significant performance degradation.

\subsubsection{Large Scale Simulations}
\begin{table}
	\centering
	\resizebox{0.95\linewidth}{!}{%
		\begin{tabular}{| c  || c || c || c || c || c |}
			\hline
			\begin{tabular}[c]{@{}l@{}} Exp.\\ NO. \end{tabular} & 
			Topology & 
			{\begin{tabular}[c]{@{}l@{}} Fabric\\~~load \end{tabular}} & 
			{Workload} & 
			{\begin{tabular}[c]{@{}l@{}} ~~\# VMs\\ per tenant \end{tabular}} & 
			{\begin{tabular}[c]{@{}l@{}} ~~Average\\ VTN scale \end{tabular}}  \\
			\hline
			1 & $k{=}8$ FT & * & enterprise & 20 & 4 \\
			\hline
			2 & $k{=}8$ FT & $3{:}1$ & * & 20 & 4 \\
			\hline
			3 & * & $3{:}1$ & enterprise & 20 & 4 \\
			\hline
			4 & $k{=}8$ FT & $3{:}1$ & enterprise & 20 & * \\
			\hline
			5 & $k{=}8$ FT & $3{:}1$ & enterprise & * & 4 \\
			\hline
	\end{tabular}}
	\caption{Experimental settings. 
		``*'' refers to that we vary the factor to isolate its impact in the experiment. 
	}\label{tab:setting}
\end{table}

In order to investigate how \sys's performance is affected by various
factors that are not covered by our small-scale testbed experiments, 
we perform detailed simulations using large scale 
datacenter topologies and empirical workloads obtained 
from production datacenters (Figure \ref{fig:flow}). 
As summarized in Table~\ref{tab:setting},
we thoroughly evaluate the performance of \sys under the 
impact of five factors: the topology, the
network/fabric load, the traffic workload, the average number of VMs 
occupied by a tenant, and the average VTN scale.

\noindent\textbf{Various Fabric Loads (Exp NO. 1):} 
In multi-tenant datacenters, fabric load can be quantified as the VM over-subscription ratio, which is defined as 
the ratio of worst-case achievable aggregate bandwidth among VM pairs 
to the total capacity of the topology. A fat-tree fabric 
has an hypervisor over-subscription of $1{:}1$ because all hypervisors may potentially send at 
the full bandwidth of their network interfaces. Thus, if the average number of 
VMs hosted by a hypervisor is $N$, the VM over-subscription ratio is $N{:}1$. 
Hereafter, we use the VM over-subscription ratio as \emph{load degree}.

We tune the datacenter load degree in the range of $1{:}1$-$7{:}1$ by varying the
number of embedded tenants. For each load, we 
create $5$ different snapshots to avoid bias. 
For each snapshot, we randomly pick a tenant to perform routing update 
using all four solutions: \sys, the local solution, ECMP and the bottomline solution. 
Thus, we perform $4$ independent experiments for the tenant. 
In each solution, after the desired routing is finalized, 
we assign each virtual link a randomly sampled flow size 
to generate traffic and compute the average FCT among all flow transfers. 
Among all $4$ experiments for the same tenant, we use the same set of flow sizes.  
After finishing the current tenant, we recover the network snapshot and re-sample 
another tenant to continue evaluation. In total, 100 routing updates are performed 
for each snapshot. 

Figure \ref{fig:baseline} plots the min-median-max distribution of the
averaged FCT across all $5$ snapshots for each load degree. 
As illustrated in the Figure \ref{fig:baseline:a}, \sys significantly outperforms 
ECMP and bottomline: up to 80\% FCT reduction for small fabric loads and at least $40$\% 
FCT reduction for all loads. Again, this is because ECMP and bottomline are static
solutions that are unaware of the network utilization. Further, 
as load degree increases, \sys offers less FCT reductions since 
there is less routing optimization space in heavily utilized network. 
Since production datacenters are typically over-provisioned, 
we can expect large performance benefits offered by \sys in production datacenters.

Due to the lack of exact mapping between virtual links and physical links, 
the local solution often produces sub-optimal routings. As shown in Figure \ref{fig:baseline:b},  
with high probabilities (up to ${\sim}60$\%), the local solution returns an overlay 
different from the one produced by \sys. Consequently, these sub-optimal 
routings result in $1.2{-}4$x FCT inflation, compared with 
the routings produced by \sys (Figure \ref{fig:baseline:a}).

\noindent\textbf{Datacenter Traffic Workload (Exp NO. 2)}: 
In Figures \ref{fig:baseline:a} and \ref{fig:baseline:b}, we use the enterprise datacenter workload. We repeat the 
same experiment using the data-mining workload to learn the impact of traffic workload on \sys's performance. 
The results, plotted in Figures \ref{fig:baseline:c} and \ref{fig:baseline:d}, show that \sys provide similar benefits for 
data-mining workload. 

\noindent\textbf{Fabric topologies (Exp NO. 3):} 
We investigate four datacenter topologies: three organized topologies (Clos, $k{=}8$ and $k{=}16$
fat-tree) and another organized topology added with random short-cuts. 
The Clos topology has the same number of hypervisors as the $k{=}8$ fat-tree topology except that the 
over-subscription ratio is $2{:}1$. All three organized topologies have different 
routing redundancy. In particular, they have $8$, $16$ and $64$ shortest paths 
between two randomly chosen hypervisors in different pods, respectively. 
The short-cut topology is built 
by adding random links into the $k{=}8$ fat-tree topology. The links are randomly 
added between ToR switches to inter-connect different pods so that 
besides traversing through the core switches, 
the inter-pod communication can alternatively use the short-cut bridges 
as well. We set the load degree as $3{:}1$ in all these topologies.  

\begin{figure}[t]
  \centering
  \mbox{
    \subfigure{\includegraphics[scale=0.55]{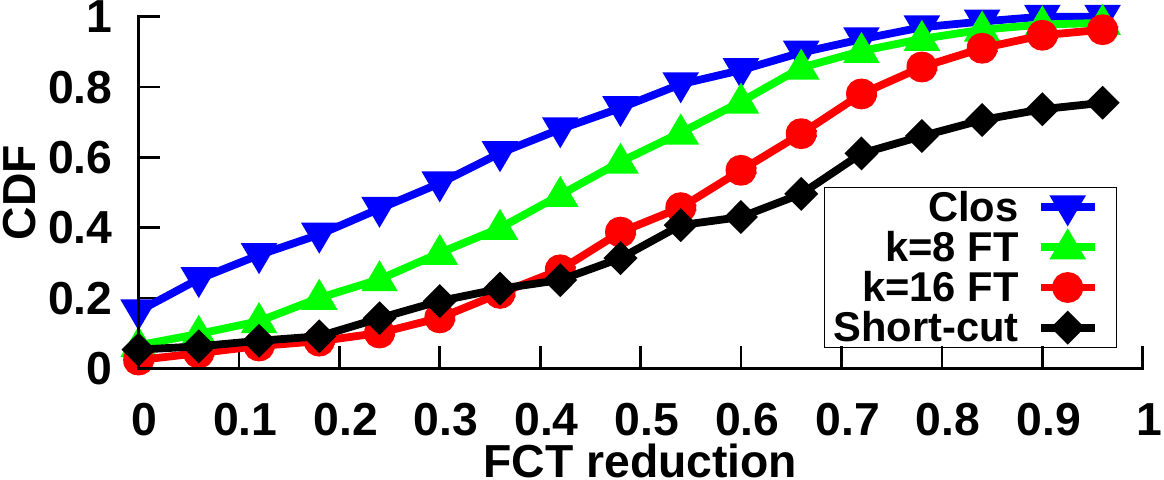}}
    }
  \caption{Impact of datacenter topology on \sys's performance of FCT reduction.
  }
  \label{fig:topology}
\end{figure}

For each topology, we perform the similar experiments as in evaluating the
impact of fabric loads. Figure \ref{fig:topology} plots the CDF of all obtained FCT 
reductions (compared with the bottomline solution). 
By comparing the results of all 3 fat-tree topologies, we conclude  
that \sys offers more benefits for topologies built with more redundant links.  
This is because \sys has larger optimization space in more redundant topologies. 

We further investigate how topology short-cuts may affect the performance of \sys. 
In organized datacenter topologies, alternative routing options typically have 
less diversity in the sense that they all have the same number of hops, 
although they are varying in terms of utilization. In contrast, short-cut topology can 
create more diverse routings with different numbers of hops as well as different utilization. 
Thus, \sys has even larger optimization space in the short-cut topology 
so as to produce more performance benefits. 

\begin{figure}[t]
  \centering
  \mbox{
    \subfigure[\label{fig:vtn_scale}Average VTN scale (Exp NO. 4)]{\includegraphics[scale=0.38]{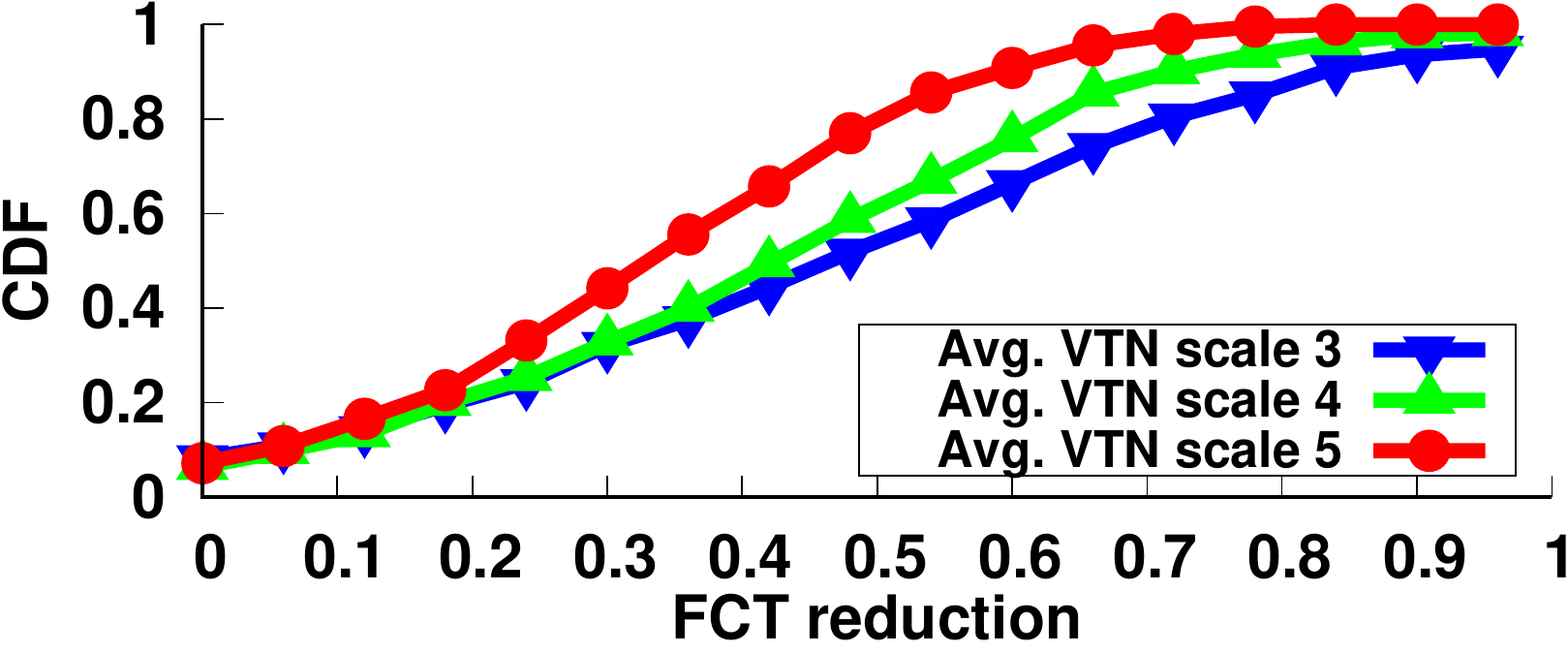}}
}
\mbox{
    \subfigure[\label{fig:vm_size}Average per-tenant VM count (Exp NO. 5)]{\includegraphics[scale=0.38]{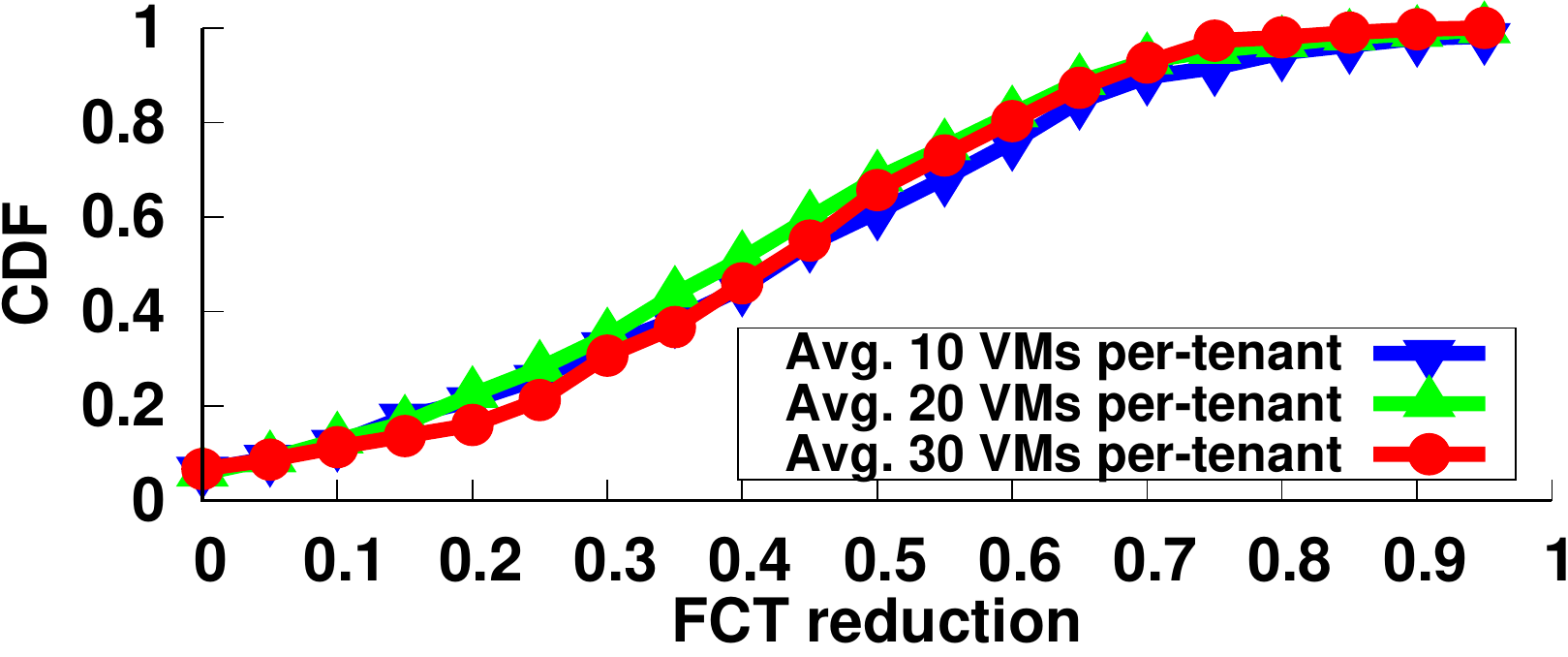}}
  }
  \caption{\sys offers consistent benefits for different VTN scales and different per-tenant VM counts.}
  \label{fig:4and5}
\end{figure}

We also investigate the average VTN scale $\mathcal{N}_h$ (the average number of hypervisors 
used by a tenant's VMs) and the average per-tenant VM count $\mathcal{N}_v$. 
We consider $\mathcal{N}_h{=}[3,4,5]$ and $\mathcal{N}_v{=}[10,20,30]$ 
since they are the common cases in our production datacenters. 
We find that \sys consistently provides performance benefits in these settings 
(Figure~\ref{fig:4and5}).

\subsection{System Properties}\label{sec:system_evaluation}

\subsubsection{Search Cost Reduction}\label{sec:search_cost_reduction}
\begin{figure}[t]
	\centering
	\mbox{
		\subfigure[\label{fig:seach_cost_reduction} Search cost reduction]{\includegraphics[scale=0.23]{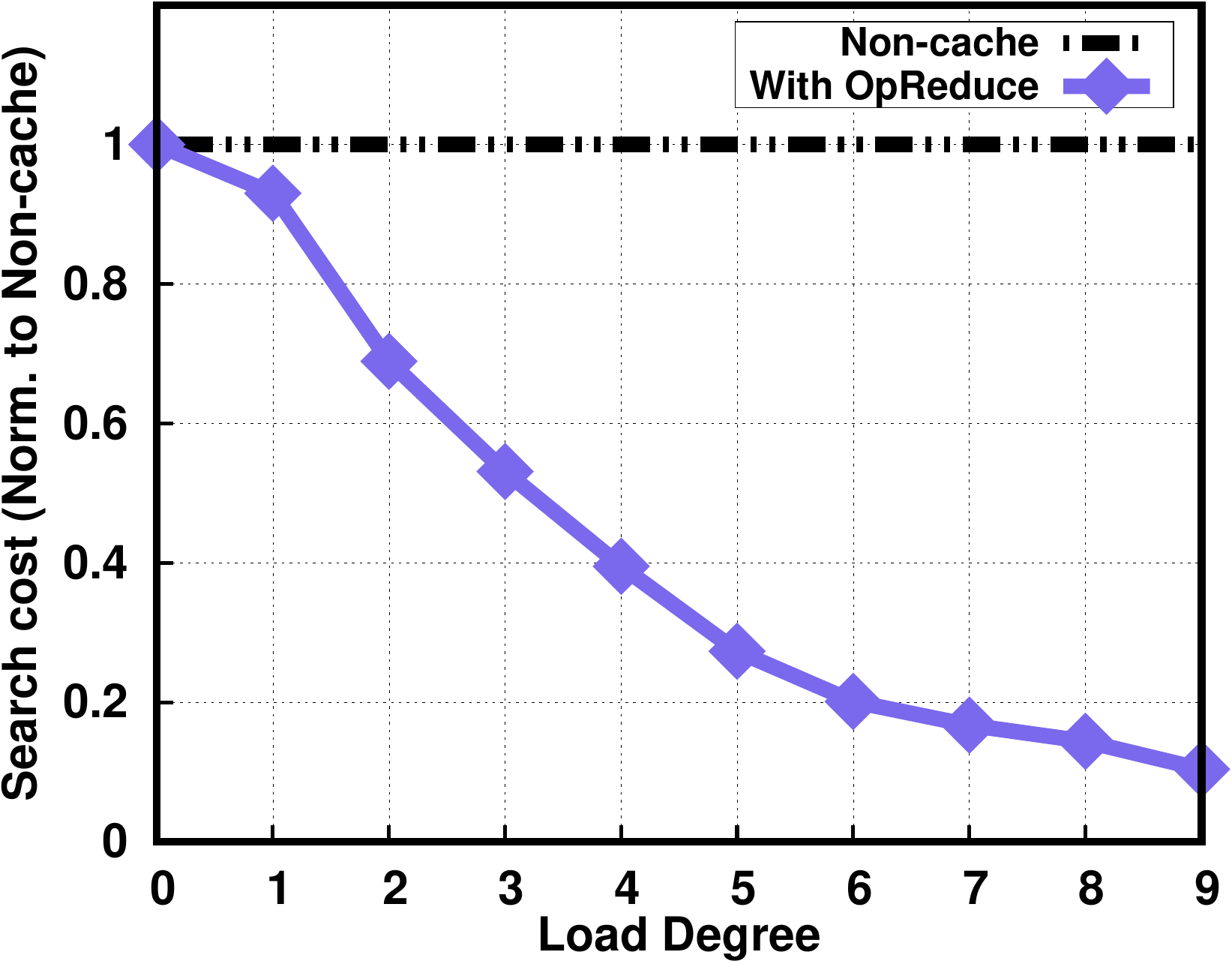}}
		\subfigure[\label{fig:catch_hit_ratio} Cache hit ratio]{\includegraphics[scale=0.23]{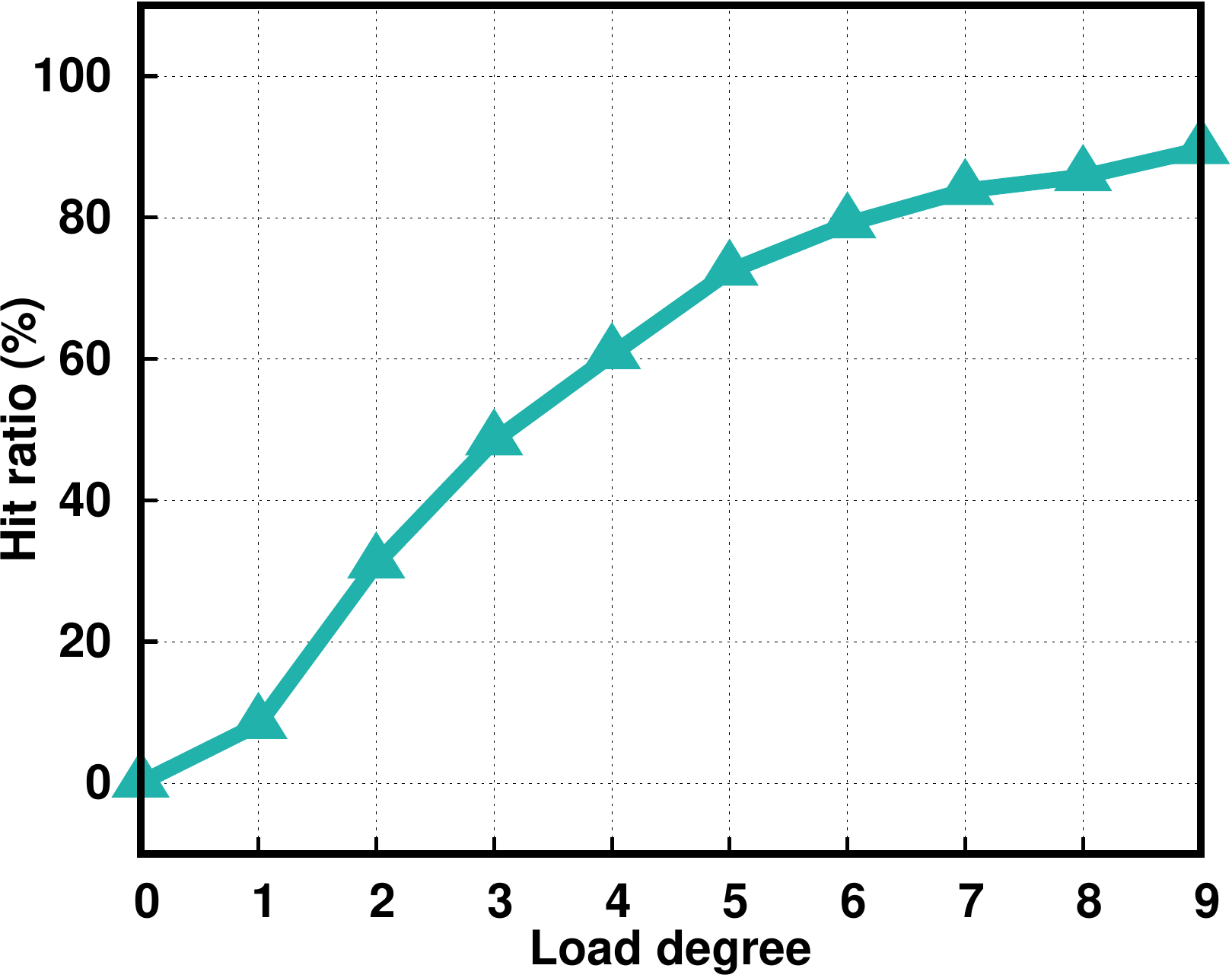}}
	}
	\caption{In spite of the numerous routing updates as load degree increases, 
		\sys introduces small search cost since the cache hit ratio increases dramatically.  
	}
	\label{fig:search_cost}
\end{figure}

The number of routing update requests is affected by the amount of embedded 
tenants, \ie the load degree. Figure~\ref{fig:seach_cost_reduction} plots search 
cost reduction under various loads for $k{=}16$ fat-tree topology. 
Although the absolute number of routing updates increase 
as load degree increases, the normalized search cost actually reduces. 
This is because \sys's knowledge base about routing candidates for 
various VM locations also expands as the load degree increases. 
As a result, the cache hit ratio dramatically increases as well (Figure~\ref{fig:catch_hit_ratio}). 
Thus, in spite of the numerous routing updates, \sys introduces small topology search cost. 

\subsubsection{Routing Cache Size}\label{sec:evaluation:cache_size}
The routing cache size is affected by two factors: the topology and the average VTN scale.  
The topology affects the cache size because it affects the number of routing candidates 
for one VM placement. The VTN scale affects the cache size since it determines 
the number of physical links in one routing candidate. 

\begin{figure}[t]
	\centering
	\mbox{
		\subfigure[\label{fig:cache_size_diff_VTN} Various VTN scales]{\includegraphics[scale=0.23]{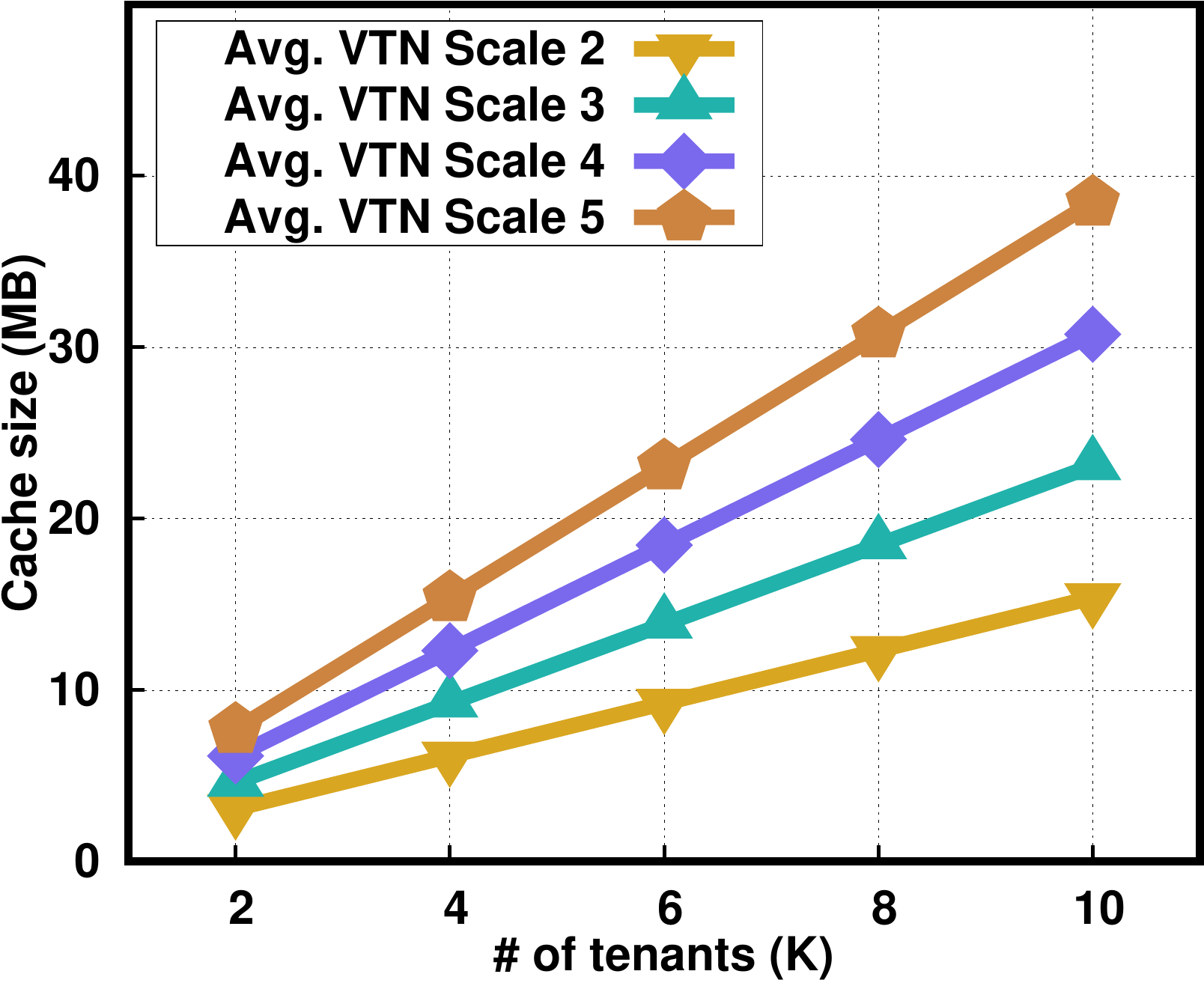}}
		\subfigure[\label{fig:cache_size_diff_topo} Various topologies]{\includegraphics[scale=0.235]{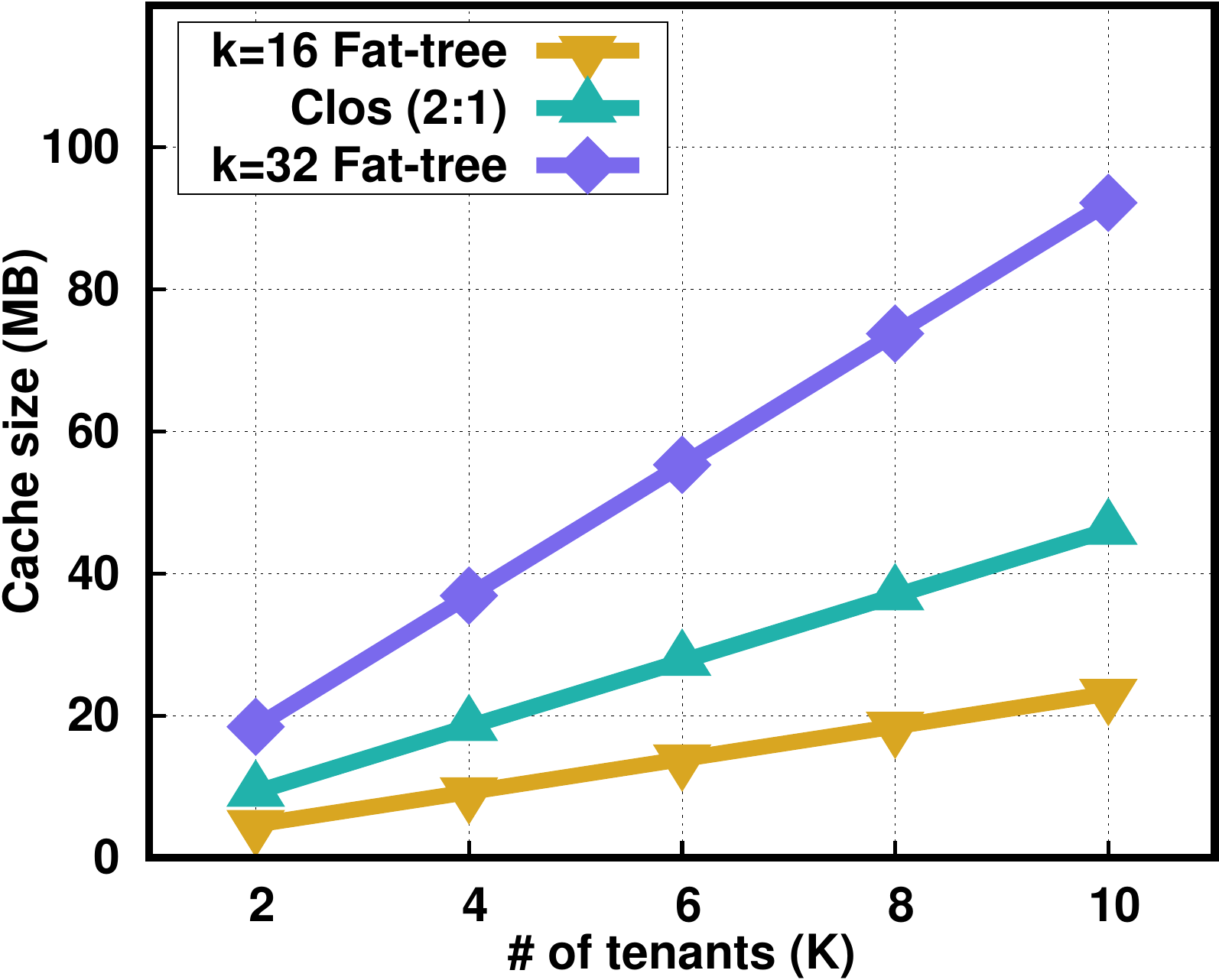}}
	}
	\caption{The routing cache size under various settings.}
	\label{fig:cache_size}
\end{figure}

Figure~\ref{fig:cache_size_diff_VTN} plots cache sizes with respect to the number of 
tenants and the average VTN scales in $k{=}16$ fat-tree topology. 
The cache size increases with the average VTN scale and linearly grows with the number of tenants. 
However, even with very sporadically distributed VTNs, \ie large average VTN scales, 
caching routing candidates for $10$K tenants consumes no more than $40$MB memory. 
Even if we consider a much larger $k{=}32$ fat-tree datacenter with over 
$8$K servers, the cache size for $10$K tenants is about hundreds of Megabytes 
(Figure~\ref{fig:cache_size_diff_topo}), 
which can be easily managed by commodity servers with gigabytes of memory. 

\subsubsection{Switch Configuration Latency}\label{sec:evaluation:switch_config}
Both SDN switches and legacy switches may co-exist in today's datacenters~\cite{hybnet}. 
Configurations on OpenFlow Switches can be finished almost 
in real time via SDN controllers such as OpenDayLight~\cite{opendaylight}, but it takes 
non-trivial time to configure legacy switches. 
Thus, to ensure that routing enforcement does not become 
the bottleneck for routing update, \sys designs a network 
action container to properly aggregate configuration tasks so as 
to reduce the overall configuration delay.

\begin{figure}[t]
	\centering
	\mbox{
		\subfigure{\includegraphics[scale=0.53]{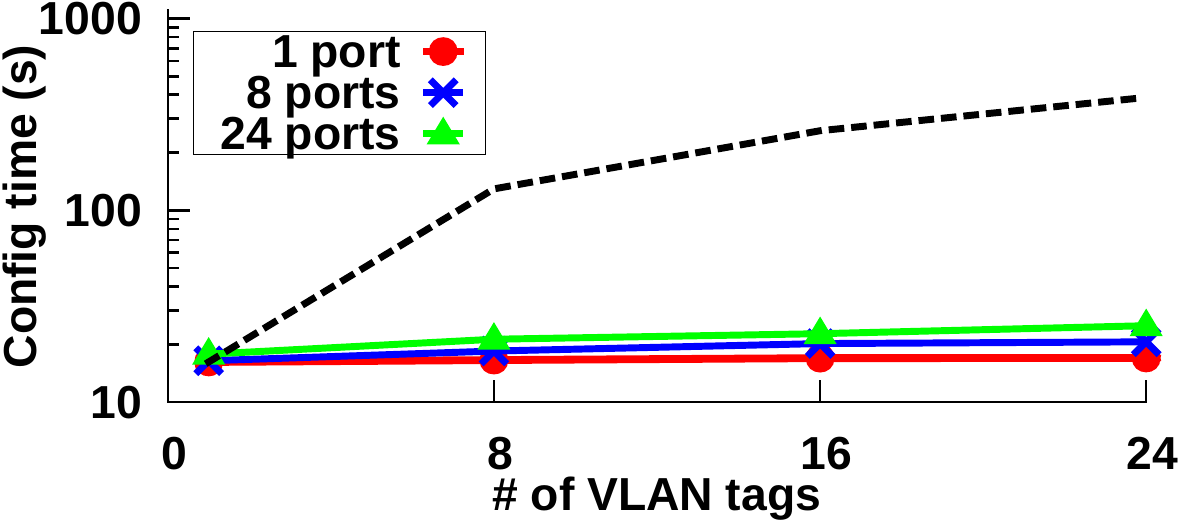}}
	}
	\caption{Delay measurements for configuring legacy switches. 
		The dash line indicates the configuration time without using 
		the network action container.
	}
	\label{fig:config_delay}
\end{figure}

In \sys's prototype, one switch configuration task is about associating one VLAN tag on 
a certain port of a switch. We notice that configuring a single port on a legacy switch 
takes almost the same amount of time as configuring multiple ports on the switch. 
Thus, our network action container aggregates all 
configuration tasks on the same switch together to perform batch configuration so as 
to reduce the overall configuration delay. Meanwhile, 
batch configurations for different switches are executed simultaneously via multi-threading. 
Figure~\ref{fig:config_delay} plots the measured configuration delay on our testbed. 
The results show that even if we simultaneously configure $24$ ports on 
one switch and on each port we configure $24$ 
VLAN tags ($576$ single configuration tasks), the overall configuration time is less than $1.5$ times the 
delay for configuring just one VLAN tag on a single port. 
Thus, with the network action container, routing enforcement can be finished timely
so that it will not be a bottleneck in practice. 

\section{Related Work and Discussion} \label{sec:related}
\parab{Multi-tenancy management.} 
Prior designs for multi-tenant datacenters, 
such as NVP~\cite{vmware} and Netlord~\cite{netlord}, focus on multi-tenancy management 
at hypervisors. For instance, NVP maintains virtual switches on each hypervisor and  
leverages a set of tunnels between each pair of hypervisors to deliver traffic for tenant VMs. 
The actual tenant traffic forwarding in the physical network is not managed. 
Several prior works have considered to perform one-time in-network routing management 
to achieve various goals, such as guaranteed bandwidth \cite{secondnet,predictable,tag}, 
bounded latency \cite{silo,qjump} and user/service isolation \cite{FlowVisor}. 
\sys, instead, focuses on managing \emph{recurrent} routing updates in our production datacenters. 

\parab{A wide variety of performance enhancement.} 
Many approaches have been proposed to improve datacenter networking 
performance. For instance, load balancing approaches \cite{conga, microte, hedera, flare}, 
priority queuing approaches \cite{pdq, pfabric, detail}, 
deadline-aware approaches~\cite{deadline} and DCTCP \cite{dctcp}
are proposed to improve latency performance. Portland~\cite{portland} and fat-tree~\cite{fat-tree} 
propose scalable datacenter architectures to support high bandwidth between servers whereas 
VL2~\cite{vl2} virtualizes datacenters into server pools to allow applications to obtain high throughput.  
Although \sys is not proposed to explicitly improve certain performance, its efficient 
tenant routing management and decoupled design allow network operators to enhance 
a wide variety of \emph{customer-interested} performance metrics, and some of these  
metrics cannot be optimized using prior approaches. 

\parab{Achieving agile routing updates.} To be readily deployable, 
\sys is augmented by SDN to perform agile in-network routing updates. B$4$~\cite{b4} and 
SWAN~\cite{swan} also adopt SDN to perform traffic engineering in wide area 
networks to achieve high inter-datacenter throughput.

\section{Conclusion}
\label{sec:conclude}
In this paper, we present \sys, a system for managing virtual tenant 
network update in multi-tenant datacenters. 
Conventional solutions that rely on topology search coupled with an objective function 
to find desired routings have at least two shortcomings: scalability 
issue for handling recurrent routing updates and the inefficiency 
for satisfying various routing requirements. To address these issues, 
\sys proposes a novel search and optimization decoupled design, which enables 
routing search result reuse and guaranteed routing optimality. We implement 
a prototype of \sys and perform extensive evaluations to valid \sys's design goals. 
Evaluation results show that \first Even for complex VTN embedding goals, 
\sys ensures routing optimality which yields significant networking performance improvement 
over conventional approaches; \second \sys greatly reduces search cost for managing 
numerous routing updates and imposes small system overhead.

\bibliography{paper}{}
\bibliographystyle{IEEEtran}

\newpage
\clearpage

\end{document}